\documentclass[11pt,a4paper]{article}
\usepackage[left=2cm ,top=25mm,bottom=25mm,right=2cm]{geometry}
\usepackage{tikz}
\usepackage[utf8]{inputenc}
\usepackage{multicol}
\usepackage{amsmath,amsthm,amssymb}
\usepackage{multicol}
\usepackage{mathtools}
\usepackage{graphicx}
\usepackage[table]{xcolor}
\usepackage{float}
\usepackage{graphicx} 
 \usepackage{multirow} 
\usepackage{hyperref}

\usepackage{subfigure}
\usepackage{mathtools}
\usepackage{pdfpages}
\numberwithin{equation}{section}

\usepackage{orcidlink}
\usepackage{cite}

\begin{document}

\thispagestyle{empty}
\begin{titlepage}
	\vspace*{0.7cm}
	\begin{center}
		{\Large {\bf Thermal effects and finite-temperature cosmology in perturbatively stabilized large volume scenarios}}
		\\[12mm]
		Vasileios Basiouris~$^{\orcidlink{0000-0002-9266-7851}}$~\footnote{E-mail: \texttt{v.basiouris@gmail.com}}
	\end{center}
	\vspace*{0.50cm}
	\centerline{\it
		Physics Department, University of Ioannina}
	\centerline{\it 45110, Ioannina, Greece}
	\vspace*{1.20cm}
	
\begin{abstract}
We analyze finite-temperature effects in the perturbative Large Volume Scenario (LVS), examining the dynamics of thermalized K\"ahler moduli and the stability of thermally induced vacua. We determine the maximum decompactification temperature $T_{\max}$, showing its dependence on winding-loop corrections in the effective theory. Moduli stabilization arises from perturbative logarithmic loop corrections to the K\"ahler potential, with higher-derivative $F^4$ terms used to test vacuum robustness. We derive cosmological implications, including bounds on reheating, and show that the model favors high scale inflationary scenarios with the heavy modulus decaying before dominating the energy density of the universe. We also analyze thermal metastability, demonstrating that the recovery of the $T=0$ vacuum is sensitive to the post-inflationary thermal history, potentially yielding either an AdS vacuum, a metastable phase, or a stable dS configuration. We conclude by briefly discussing the relevance of finite-temperature effects for entropy-assisted tunneling and the broader string landscape.
\end{abstract}
\end{titlepage}

\section{Introduction}
A long-standing problem of string phenomenology is the determination of dynamics of the uncharged massless fields, called moduli. Over the decades, significant advances have been made in tackling this problem. In type IIA string theory, turning on background fluxes can fix all geometric moduli \cite{Derendinger:2004jn, Villadoro:2005cu, DeWolfe:2005uu, Giddings:2001yu}, despite the fact that the mathematical study of fluxes is much more complicated compared to type IIB theory. Type IIB compactifications require additional contributions from both perturbative and non-perturbative effects \cite{hep-th/0301240, hep-th/0502058}. In comparison, this scheme allows for a large set of solutions where flux backreaction is mostly captured by a warp factor, making moduli stabilization better understood in IIB Calabi-Yau flux compactifications. A prominent example of successful IIB moduli stabilization is the Large Volume Scenario (LVS) introduced in \cite{hep-th/0502058}, which has applications in both cosmology and particle physics. In this framework \cite{arXiv:0805.1029}, $\alpha^{\prime}$ and $g_s$ corrections combine together with non-perturbative effects to stabilize the K{\"a}hler moduli, while background fluxes stabilize the dilaton and complex structure moduli. Unlike the KKLT approach \cite{hep-th/0301240}, LVS does not require fine-tuning of flux values, and the Calabi-Yau volume is stabilized on an exponentially large scale. This setup provides a trustworthy four-dimensional effective theory and allows for creation of hierarchies relevant for phenomenology.\par
Recently, several arguments have questioned the validity of constructing de Sitter vacua within a 4D effective supergravity framework. These concerns have contributed to the formulation of the swampland de~Sitter conjecture~\cite{Vafa:2005ui, Ooguri:2006in, Grana:2021zvf}. Both of the aforementioned approaches in constructing stable vacua admit an AdS vacuum, while a dS minimum emerges with a supersymmetric minimum at infinity after the uplifting. These two vacua are separated by a barrier $V_b$, whose magnitude is specified by the scale of the AdS point. Most swampland related criticisms focus on the reliability of non-perturbative effects, the role of uplift mechanisms involving anti-D3 branes, and the precision of quantum corrections. The modulus controlling the overall volume of the Calabi-Yau (CY) manifold interacts with all sources of energy due to the Weyl rescaling required to obtain the four-dimensional supergravity effective action in the Einstein frame. Consequently, any energy contribution that exceeds the height of the potential barrier can push the system toward decompactification. For example, during inflation, the energy of the inflaton may generate an additional uplift term, potentially causing an infinite volume runaway \cite{hep-th/0411011}. Similarly, after inflation, inflaton decay produces a high-temperature thermal plasma, leading to temperature-dependent corrections to the moduli potential, which could destabilize the moduli if they exhibit runaway behavior. The temperature at which the finite-temperature effects \cite{Binetruy:1984yx, Binetruy:1984wy, Buchmuller:2004tz, Anguelova:2007at, Fischler:2006xh, Gross:1982cv} dominate over the zero-temperature potential, i.e., the decompactification temperature, is approximately of order $T_{\rm max} \sim V_b^{1/4}$, with $V_b$ denoting the height of the potential, and sets an upper bound on the reheating temperature. On the other hand, if the finite-temperature potential develops new minima rather than running away, the system may undergo phase transitions \cite{Linde:1978px, Linde:1980tt, Linde:1981zj, Linde:1982tg}, potentially affecting early universe evolution and leaving observable imprints today. High-temperature minima could also influence the likelihood of the universe settling in a metastable state at $T=0$. Studies of several toy models \cite{hep-th/0611006, hep-th/0610334, hep-th/0611018, hep-th/0702168, Papineau:2008xf, hep-th/0404168} suggest that despite the existence of a supersymmetric global minimum, systems starting in a high-temperature minimum often end up in a long-lived metastable state with broken supersymmetry at low temperatures. If analogous behavior occurs in realistic scenarios, it could provide conceptual insights into the present accelerated expansion of the universe.\par
In the current work, we examine the aforementioned questions under the perspective of perturbative moduli stabilization scheme \cite{Leontaris:2022rzj, Basiouris:2020jgp, Basiouris:2021sdf, Bera:2024zsk, Hai:2025wvs, Chakraborty:2025wqn} considering also higher derivative corrections and winding loop corrections. This combination results in a distinct hierarchy in the masses and couplings of the corresponding moduli, leading to thermalization of a direction transverse to the volume modulus as it is summarized in Section 2. In Section 3. and Section 4., the finite temperature effects augment the effective theory and their effects is to shift the dS vacuum to a new point in the moduli space, where we manage to provide analytic expression both for this dynamical phenomenon as well as the upper bound on temperature. To our knowledge, this is the first approach in the literature that unravels the association of $T_{max}$ with winding loop effects and the uplifting mechanism, while we provide analytical evidence that the thermalized field decays before its energy density starts to overclose the universe. Finally in Section 5., the potential thermal metastability of the underlying theory is discussed and a critical temperature is indicated, which signifies that Hawking-Moss (HM) \cite{Hawking:1981fz} instanton solutions along the volume direction do not destabilize the thermal vacuum. Conversely, during the cooling of the universe, the reheating scale above this critical temperature could either induce a metastable dS augmented by an AdS vacuum-suggesting a possible first-order phase transition-or lead to a stable $T=0$ vacuum via a continuous second-order–like transition. Finally, we offer some remarks on the plausible association between finite temperature effects and quantum–entropy–assisted phase transitions, highlighting their relevance with the fate metastable vacua of the string landscape.

\section{The structure of the model}
In this paper, we consider an effective theory model within the type IIB string framework, where the dilaton and the complex structure moduli are stabilized by turning on 3-from fluxes. The K{\"a}hler potential receives corrections either from perturbative dynamics or from non-perturbative corrections in the superpotential. The well-known structures of KKLT and LVS vacua incorporate non-perturbative corrections as the primary mechanism to stabilize the K{\"a}hler sector. The non-scale structure of effective theory is broken $\alpha^{\prime}$ effects or considering $g_s$ corrections in the K{\"a}hler potential. Apart from the BBHL $\alpha^{\prime 3}$ perturbative corrections \cite{hep-th/0204254}, we currently consider 1-loop logarithmic effects scaling as $\sim \eta \log(\tau)$, stemming from intersecting $D_7$ branes \cite{Antoniadis:2018hqy, Antoniadis:2019rkh}, where the final form of the K{\"a}hler potential in terms of the 4-cycle moduli is written as:
\begin{equation}
    \mathcal{K} = -2 \log \left(\sqrt{\tau _1 \tau _2 \tau _3} + \xi + \sum_i\left(\eta_i  \log \left(\tau _i\right)\right)\right)~.
\end{equation}
The form of the compactified space considered here is toroidal looking volume form and it follows from the proposal of \cite{Antoniadis:2018hqy, Basiouris:2020jgp}, and the superpotential receives no non-perturbative corrections \cite{hep-th/9906070}.
\begin{equation}
    \mathcal{V}=\sqrt{\tau_1 \tau_2 \tau_3}~.
\end{equation}
\begin{equation}
    \mathcal{W} =\mathcal{W}_0~.
\end{equation}
To validate the above proposal of the compactified space, one could revisit the Kreuzer-Skarke \cite{Kreuzer:2000xy, Altman:2014bfa} database of CY's manifolds, where the three moduli case could descend from a manifold with $h^{1,1}=3$. As pointed out in the analysis performed in \cite{Leontaris:2022rzj}, a manifold with the following toric data could potentially result into the proposed volume form of the current model. One such CY threefold corresponding to the polytope Id: 249 in the CY database of \cite{Altman:2014bfa} can be defined by the following toric data:
\begin{center}
\begin{tabular}{|c|ccccccc|}
\hline
Hyp & $x_1$  & $x_2$  & $x_3$  & $x_4$  & $x_5$ & $x_6$  & $x_7$       \\
\hline
4 & 0  & 0 & 1 & 1 & 0 & 0  & 2   \\
4 & 0  & 1 & 0 & 0 & 1 & 0  & 2   \\
4 & 1  & 0 & 0 & 0 & 0 & 1  & 2   \\   
\hline
& $K3$  & $K3$ & $K3$ &  $K3$ & $K3$ & $K3$  &  SD  \\
\hline
\end{tabular}
\end{center}
The Hodge numbers of the threefold are $(h^{2,1}, h^{1,1}) = (115, 3)$, the Euler number is defined by $\chi=-224$, where the SR ideal is written as:
\begin{equation}
{\rm SR} =  \{x_1 x_6, \, x_2 x_5, \, x_3 x_4 x_7 \} \,. 
\end{equation}
The structure of the divisor topologies has been analyzed with {\it cohomCalg} \cite{Blumenhagen:2010pv, Blumenhagen:2011xn}, where this framework can be represented by the following Hodge diamonds:
\begin{align}
K3 \equiv \begin{tabular}{ccccc}
    & & 1 & & \\
   & 0 & & 0 & \\
  1 & & 20 & & 1 \\
   & 0 & & 0 & \\
    & & 1 & & \\
  \end{tabular}, \qquad \quad {\rm SD} \equiv \begin{tabular}{ccccc}
    & & 1 & & \\
   & 0 & & 0 & \\
  27 & & 184 & & 27 \\
   & 0 & & 0 & \\
    & & 1 & & \\
  \end{tabular}.
\end{align}
The basis of smooth divisors $\{D_1, D_2, D_3\}$ define the following intersection polynomial, which admit only one non-zero triple intersection number:
\begin{equation}
I_3 = 2\, D_1\, D_2\, D_3,
\end{equation}
while the second Chern-class of the CY is given by,
\begin{equation}
c_2(CY) = 5 D_3^2+12 D_1 D_2 + 12 D_2 D_3+12 D_1 D_3.
\end{equation}
Afterwards, the K\"ahler form can be derived through $J = \sum_{\alpha =1}^3 t^\alpha D_\alpha$, where the overall volume and the 4-cycle volume moduli follow from:
\begin{equation}
\mathcal{V} = 2\, t^1\, t^2\, t^3, \quad \quad \tau_1 = 2\, t^2 t^3,  \quad  \tau_2 = 2\, t^1 t^3, \quad  \tau_3 = 2 \,t^1 t^2 \,.
\label{Taus}
\end{equation}
Finally, the final form of the compactified volume could be expressed by the combination of the two-cycles as:
\begin{equation}
\mathcal{V} = 2 \, t^1\, t^2\, t^3 = t^1 \tau_1 = t^2 \tau_2 = t^3 \tau_3 = \frac{1}{\sqrt{2}}\,\sqrt{\tau_1  \tau_2 \tau_3}~.
\end{equation}
This validates the proposal for a toroidal like volume in the current model, where it possesses an exchange symmetry $1 \leftrightarrow 2 \leftrightarrow 3$ for the moduli descending from the exchange of the basis divisors $K3$. In the current work, we assume the simplified version of $\mathcal{V} = \sqrt{\tau_1\tau_2\tau_3}$ adapted in the aforementioned studies. Further, the K\"ahler cone for this setup is described by the conditions below,
\begin{equation}
\text{K\"ahler cone:}\;\;  t^1 > 0\,, \quad t^2 > 0\,, \quad t^3 > 0\,.
\end{equation}
The effective scalar potential can be expressed by the following form:
\begin{equation}
    V_{eff} = e^{\mathcal{K}} \bigg( \sum_{I,J} D_I \mathcal{W} \mathcal{K}^{I\bar{J}}D_{\bar{J}}\overline{\mathcal{W}}-3|W|^2\bigg)~.
\end{equation}
\begin{equation}
    V_{eff} = \frac{3 \mathcal{W}_0^2 g_s (-16 \eta +\xi +4 \eta  \log (\mathcal{V}))}{4 \mathcal{V}^3} + \mathcal{O}(\frac{1}{\mathcal{V}^4})~.
\end{equation}
Additionally, sub-leading string loop effects can be classified to two categories: the ones derived from KK-corrections and the other from winding loop effects \cite{hep-th/0508043, Cicoli:2007xp, Berg:2004ek, Berg:2005yu}. The K{\"a}hler corrections of this origin are written as:
\begin{align}
    \mathcal{K}_{g_s}^{KK} = g_s \sum_{\alpha} \dfrac{C^{KK}_{\alpha}t_{\perp}^{\alpha}}{\mathcal{V}},\quad \mathcal{K}_{g_s}^{W} = \sum_{\alpha} \dfrac{C^W_{\alpha}}{\mathcal{V}t^{\alpha}_{\cap}},
\end{align}
where the $\mathcal{C}_i$ parameters are functions that depend on the moduli of the complex structure and the dynamics of open strings, while $t_i$ define the form of 2-cycles of the underlying geometry. In particular, 2-cycle volume moduli $t_{\perp}^{\alpha}$ denote the transverse space between the $D_7$-branes and the $O_7$-planes. In contrast, $t_{\cap}^{\alpha}$ stands for the volume of the curve at the intersection point of intersecting $D_7$-branes in the background geometry. Moreover, higher derivatives corrections could also be included, whose origin is due to $\mathcal{O}(F^4)$ and their form on a given Calabi-Yau compactification are expressed as \cite{Ciupke:2015msa}:
\begin{equation}
    V_{F^4} = -\kappa^2 \dfrac{\lambda|\mathcal{W}_0|^4}{g_s^{3/2}\mathcal{V}^4}\Pi_i t^i, \; \kappa = \dfrac{g_s}{8\pi},
\end{equation}
where $\lambda$ is a constant with no moduli dependence and $\Pi_i$ are topological numbers defined by the harmonic 2-form of the compactification. In the current setup, we assume the existence of winding loop corrections along with higher derivative corrections, where the KK effects can be neglected since they parameterize the distance among the stacks of $D_7$-branes and $O_7$-planes, and it can be assumed that $C^{KK}\ll 1$. Then, the contribution to the scalar potential arising from the aforementioned sub-leading effects is given as:
\begin{align}
    V_{F^4} = -\frac{3 \lambda  \left(t_1+t_2+t_3\right) \mathcal{W}_0^4 \sqrt{g_s}}{8 \pi ^2 \mathcal{V}^4}, \quad V_w = -\frac{\mathcal{W}_0^2 g_s\left(\frac{\mathcal{C}_1}{t_1}+\frac{\mathcal{C}_2}{t_2}+\frac{\mathcal{C}_3}{t_3}\right)}{8\pi \mathcal{V}^3},
\end{align}
\begin{equation}
    t_1 = \sqrt{\dfrac{\tau_2 \tau_3}{2\tau_1}},\; t_2 = \sqrt{\dfrac{\tau_1 \tau_3}{2\tau_2}},\; t_3 = \sqrt{\dfrac{\tau_1 \tau_2}{2\tau_3}}~.
\end{equation}
Beyond the structure of the scalar potential at zero temperature, one-loop finite temperature effects parametrize the interactions between moduli and gauge bosons, which lead to thermalization and eventually to loop corrections of following form:
\begin{equation}
    V_{eff} = V_{eff}(T=0) + V^{loop}(T)~.
\end{equation}
These finite temperature corrections $V^{loop}(T)$ have a generic loop expansion at 1-loop and 2-loop level, which is described by \cite{Dolan:1973qd, Jackiw:1974cv, Anguelova:2007ex}:
\begin{align}
    &V^{1-loop} = \frac{T^2}{24} \left(Tr[M_{\tilde{\phi }}^2]+Tr[M_{\phi }^2]\right)-\frac{\pi ^2 T^4}{90} \left(g_b+g_f\right),\\
    &V^{2-loop} = T^4 \left(\kappa_1 g_{MSSM}^2+\kappa_2 g_{\phi XX}^2 m_{\phi }^2+\kappa_3 g_{\tilde{\phi} XX}^2 m_{\tilde{\phi} }^2\right)+...,\label{2loop}
\end{align}
where $g_b, \;g_f$ are the degrees of freedom of bosons and fermions respectively, and $M^2$ moduli-dependent bosonic and fermionic mass matrices of all the particles forming the thermal plasma. The leading order expansion of the 2-loop expansion is parametrized by contributions of loop diagrams. The terms in \eqref{2loop} descend from two loops involving MSSM particles or diagrams with one modulus and two gauge bosons $X$. The mass of the supersymmetric partners $\tilde{\phi}$ can be computed from the following formula:
\begin{equation}
    m^2_{\tilde{\phi}} = Tr[M^2_f] = e^G K^{i\bar{j}}\; K^{l\bar{m}} (\nabla_{i} G_l + \dfrac{G_i G_l}{2})(\nabla_{\bar{j}} G_{\bar{m}} + \dfrac{G_{\bar{j}} G_{\bar{m}}}{2}),
\end{equation}
where $G = \mathcal{K} + \log|\mathcal{W}|$ is the supergravity K{\"a}hler invariant potential and $\nabla_i G_j = G_{ij}-\Gamma^l_{ij}G_l$, with $\Gamma^l_{ij} = K^{l\bar{m}} \partial_i K_{j\bar{m}}$ being the geometric connection. In effective $\mathcal{N} =1$ supergravity, fermion masses arise from the second derivatives of the superpotential and the K{\"a}hler potential, while due to the LVS behavior of the vacuum, we expect that these particles are much lighter than the scalar masses. Thus, we could safely assume that $m^2_{\phi}\gg m^2_{\tilde{\phi}}$.\par
Before studying the behavior of the finite temperature effects, we first focus on the stability of the underlying theory. The minimization of the moduli can be more easily performed on the orthonormal following basis \cite{Antoniadis:2018ngr}, which can be derived through the leading order diagonal (or large volume expansion) structure of the K{\"a}hler metric.
\begin{align}
    K_{ij} \cong \begin{pmatrix}
        \dfrac{1}{4\tau_1^2} & 0 & 0\\
        0 & \dfrac{1}{4\tau_2^2} & 0 \\
        0 & 0 & \dfrac{1}{4\tau_3^2}
    \end{pmatrix} + \mathcal{O}(\eta,\xi),\; K^{-1}_{ij} \cong \begin{pmatrix}
        4\tau_1^2 & 0 & 0\\
        0 & 4\tau_2^2 & 0 \\
        0 & 0 & 4\tau_3^2
    \end{pmatrix}+ \mathcal{O}(\eta,\xi)~.
\end{align}
\begin{align}
    t_i= \dfrac{1}{\sqrt{2}}\ln (\tau_i)~.
\end{align}
Since we want to study the volume modulus $\mathcal{V}$ direction, it is better to find the definition of the volume direction and the two perpendicular directions in the normalized fields basis. An option is to choose the following directions:
\begin{align}
    &t = \dfrac{1}{\sqrt{3}}(t_1 + t_2 + t_3) =\dfrac{\sqrt{6}}{3}\ln (\mathcal{V}),\\
    &u = \dfrac{1}{\sqrt{2}} (t_1 -t_2 ),\label{vbas}\\
    &v = \dfrac{1}{\sqrt{6}} (t_1 +t_2 -2t_3)\label{tbas}~.
\end{align}
In this basis, the effective potential reads as:
\begin{align}
    V_{eff}(T=0) =&\; \frac{3}{4} e^{-3 \sqrt{\frac{3}{2}} t} \mathcal{W}_0^2 g_s \left(\xi +2 \eta  \left(\sqrt{6}t-8\right)\right) -\frac{\mathcal{W}_0^2 g_s e^{-5 \sqrt{\frac{2}{3}} t-u-\frac{2 v}{\sqrt{3}}} \left(\mathcal{C}_3e^u+e^{\sqrt{3} v} \left(\mathcal{C}_2+\mathcal{C}_1 e^{2u}\right)\right)}{8 \pi } - \notag \\
    & -\frac{3 \lambda  \mathcal{W}_0^4 \sqrt{g_s}\left(e^{u+\sqrt{3} v}+e^{2 u}+1\right) e^{-\frac{11t}{\sqrt{6}}-u-\frac{v}{\sqrt{3}}}}{8 \pi ^2}~.
\end{align}
The minimal values of the three directions can be written at leading order as:
\begin{align}
    u_{min} = \frac{1}{2} \log \left(\frac{\mathcal{C}_2}{\mathcal{C}_1}\right)+ \mathcal{O}(\lambda),\; v_{min} =\frac{\log \left(\frac{\mathcal{C}_3}{\sqrt{\mathcal{C}_1 \mathcal{C}_2}}\right)}{ \sqrt{3}}+ \mathcal{O}(\lambda), \label{uv1}\\
    t_{min} = \frac{52 \eta -3 \xi +36 \eta  \;W_{0}\left(\frac{5 (\frac{\mathcal{C}_1 \mathcal{C}_2 \mathcal{C}_3}{ \eta^3})^{1/3} e^{\frac{\xi }{12 \eta }-\frac{13}{9}}}{108 \pi }\right)}{6 \sqrt{6} \eta } + \mathcal{O}(\lambda)~.
\end{align}
The $W_{0}$ branch of the Lambert function denote the minimum of the potential along the volume direction. We can observe that the volume modulus is mainly stabilized by the perturbative corrections, while the subleading winding loop corrections stabilize the transverse directions. The higher derivative corrections scaling as $\sim \mathcal{O}(F^4)$ are added to examine the robustness of the minima against sub-leading corrections. Now, we perform the diagonalization of the mass matrix in the original basis ($\tau_1,\tau_2,\tau_3$), in order to deduce the exact hierarchy in the particle spectrum and in the couplings with the gauge bosons, concluding that way whether the normalized fields couple sufficiently strongly with the thermal bath. To work out the canonical normalization \cite{Cicoli:2010ha}, we can expand around the vev of each modulus as:
\begin{align}
    \tau_i = \langle \tau_i \rangle +\delta \tau_i,
\end{align}
where then the Lagrangian will be written as:
\begin{align}
    \mathcal{L}=\dfrac{1}{2} K_{ij} \partial_{\mu}\delta\tau_i\partial^{\mu}\delta\tau_j -V - \dfrac{1}{2}V_{ij} \tau_i \tau_j ~.
\end{align}
Writing the original moduli $\delta \tau_i$ in terms of the canonically normalized fields $\delta \phi_i$, then the Lagrangian is written as:
\begin{equation}
    \mathcal{L} = \dfrac{1}{2} \sum_i \partial_{\mu} \delta \phi_i \partial ^{\mu} \delta \phi_i - V - \dfrac{m_i^2}{2} \sum_i \delta \phi_i^2,\quad \delta \tau_i = \begin{pmatrix}
        \chi_1\\
        \chi_2\\
        \chi_3
    \end{pmatrix}\dfrac{\delta \phi_i}{\sqrt{2}},
\end{equation}
where $\chi_i$ are the components of the eigenvectors. The mass-squared matrix is computed by $M^2_{ij}= \frac{1}{2} (K^{-1}_{ik})V_{kj}$, while the eigenvectors are normalized by $\vec{\alpha}^T \cdot K \cdot \vec{\alpha} =1$. The diagonal elements of the mass matrix are given below as:
\begin{align}
    &M^2_{11} = \frac{9 \mathcal{W}_0^2 g_s (-272 \eta +15 \xi +60 \eta  \log (\mathcal{V}))}{8 \mathcal{V}^3} - \frac{\mathcal{W}_0^2 g_s \left(\tau _1 \tau _2 \left(\mathcal{C}_1 \tau _1+3 \mathcal{C}_2 \tau
   _2\right)+3 \mathcal{C}_3 \mathcal{V}^2\right)}{2 \pi  \tau _1 \tau _2 \mathcal{V}^4}-\notag\\
   &-\frac{15 \lambda  \mathcal{W}_0^4 \sqrt{g_s} \left(3 \tau _1^2 \tau _2^2+3 \tau _1 \mathcal{V}^2+7 \tau _2\mathcal{V}^2\right)}{16 \pi ^2 \tau _1 \tau _2 \mathcal{V}^5},\\
   &M^2_{22}=M^2_{11}(\mathcal{C}_1\leftrightarrow \mathcal{C}_2,\tau_1 \leftrightarrow \tau_2) ,\; M^2_{33}=M^2_{11}(\mathcal{C}_1\leftrightarrow 3\mathcal{C}_1,\mathcal{C}_3\leftrightarrow \frac{1}{3} \mathcal{C}_3)+ \frac{15 \lambda  \mathcal{W}_0^4 \sqrt{g_s} \left(\mathcal{V}^2-\tau _1^2 \tau _2\right)}{4 \pi ^2 \tau _1
   \mathcal{V}^5}~.
   \end{align}
In the above expressions, we have implicitly denoted the exchange of coefficients/moduli with $\leftrightarrow$ that is needed for the rest of the diagonal elements. The off-diagonal elements are expressed as:
\begin{align}
   &M^2_{12} = \frac{9 \tau _1 \mathcal{W}_0^2 g_s (-56 \eta +3 \xi +12 \eta  \log (\mathcal{V}))}{8 \tau _2 \mathcal{V}^3}  -\frac{\mathcal{W}_0^2 g_s \left(\tau _1 \tau _2 \left(\mathcal{C}_1 \tau _1+\mathcal{C}_2 \tau _2\right)+2\mathcal{C}_3 \mathcal{V}^2\right)}{\sqrt{2} \pi  \tau _2^2 \mathcal{V}^4}-\notag\\
   &-\frac{9 \lambda  \mathcal{W}_0^4 \sqrt{g_s} \left(3 \tau _1^2 \tau _2^2+5 \tau _1 \mathcal{V}^2+5 \tau _2
   \mathcal{V}^2\right)}{16 \pi ^2 \tau _2^2 \mathcal{V}^5},\;\; M^2_{21} = M_{12}^2(\tau_2 \leftrightarrow \tau_1)~.\\\notag \\
   &M^2_{13} = \frac{9 \tau _1^2 \tau _2 \mathcal{W}_0^2 g_s (-56 \eta +3 \xi +12 \eta  \log (\mathcal{V}))}{8\mathcal{V}^5} -\frac{\tau _1 \mathcal{W}_0^2 g_s \left(\tau _1 \tau _2 \left(\mathcal{C}_1 \tau _1+2 \mathcal{C}_2 \tau_2\right)+\mathcal{C}_3 \mathcal{V}^2\right)}{\sqrt{2} \pi  \mathcal{V}^6}-\notag\\
   &-\frac{9 \lambda  \tau _1 \mathcal{W}_0^4 \sqrt{g_s} \left(5 \tau _1^2 \tau _2^2+3 \tau _1 \mathcal{V}^2+5
   \tau _2 \mathcal{V}^2\right)}{16 \pi ^2 \mathcal{V}^7},\; M_{31}^2= M_{13}^2(g_s\mathcal{W}_0^2\tau_1 \rightarrow g_s\mathcal{W}_0^2 \frac{\tau_3^2}{\tau_1},\lambda\rightarrow \lambda \frac{\tau_3^2}{\tau_1})~.\\\notag \\
   &M^2_{23} = \frac{9 \tau _1 \tau _2^2 \mathcal{W}_0^2 g_s (-56 \eta +3 \xi +12 \eta  \log (\mathcal{V}))}{8\mathcal{V}^5} -\frac{\tau _2 \mathcal{W}_0^2 g_s \left(\tau _1 \tau _2 \left(2 \mathcal{C}_1 \tau _1+\mathcal{C}_2 \tau
   _2\right)+\mathcal{C}_3 \mathcal{V}^2\right)}{\sqrt{2} \pi  \mathcal{V}^6}-\notag \\
   &-\frac{9 \lambda  \tau _2 \mathcal{W}_0^4 \sqrt{g_s} \left(5 \tau _1^2 \tau _2^2+5 \tau _1 \mathcal{V}^2+3\tau _2 \mathcal{V}^2\right)}{16 \pi ^2 \mathcal{V}^7},\; M^2_{32} = M^2_{23}(g_s\mathcal{W}_0^2\tau_2 \rightarrow g_s\mathcal{W}_0^2 \frac{\tau_3^2}{\tau_2},\lambda\rightarrow \lambda \frac{\tau_3^2}{\tau_2})~.
\end{align}
The eigenvalues of the mass matrix can be computed using the invariants of the mass matrices, the trace and the determinant. More specifically, the characteristic polynomials and the corresponding eigenvalues are expressed by:
\begin{align}
    Tr[M^2] = \sum_i m_i^2,\quad \dfrac{Det[M^2]}{Tr[M^2]^2}\sim \dfrac{m_1^2m_2^2m_3^2}{m_1^4}\sim \dfrac{m_2^2 m_3^2}{m_1^2}~.
\end{align}
The heaviest eigenvalue ($m_1^2\gg m_{2,3}^2$) is effectively parametrized by the scale of the perturbative corrections and it is given as:
\begin{align}
    &m_1^2 \cong Tr[M^2]\cong\frac{9}{8} e^{-3 \sqrt{\frac{3}{2}} t} \mathcal{W}_0^2 g_s \left(-272 \eta +15 \xi +30 \sqrt{6} \eta 
   t\right) -\frac{7 \mathcal{W}_0^2 g_s e^{-5 \sqrt{\frac{2}{3}} t-u-\frac{2 v}{\sqrt{3}}} \left(\mathcal{C}_3e^u+e^{\sqrt{3} v} \left(\mathcal{C}_2+\mathcal{C}_1 e^{2 u}\right)\right)}{2 \pi }~.\label{massphi}
\end{align}
The two smallest eigenvalues, given the large hierarchy, can be given by the algebraic expressions of the $2\times2$ submatrix, where the trace of the $M^2$ matrix specifies the quantities as:
\begin{align}
    &m^2_{2,3} \cong \frac{1}{2} (Tr[M^2]-\frac{1}{2}(Tr[M^2]^2-Tr[M^4])^{1/2}),\;\; A=(Tr[M^2]^2-Tr[M^4])^{1/2}~.\\
    &A=\frac{1}{64} e^{-3 \sqrt{6} t} \mathcal{W}_0^4 g_s^2 \left(17649 \xi ^2+8 \eta ^2 \left(52947 t^2-160104 \sqrt{6} t+726208\right)+12\eta  \xi  \left(5883 \sqrt{6} t-53368\right)\right) \notag \\
    &-\frac{117 \lambda  \left(e^{2 u}+1\right) \mathcal{W}_0^6 g_s^{3/2} \left(-2368 \eta +129 \xi +258 \sqrt{6} \eta  t\right) e^{-10\sqrt{\frac{2}{3}} t-u-\frac{v}{\sqrt{3}}}}{64 \pi ^2}\notag \\
    & -\frac{63 \mathcal{W}_0^4 g_s^2 \left(-272 \eta +15 \xi +30 \sqrt{6} \eta  t\right) e^{-\frac{19 t}{\sqrt{6}}-u-\frac{2 v}{\sqrt{3}}}\left(\mathcal{C}_3 e^u+e^{\sqrt{3} v} \left(\mathcal{C}_2+\mathcal{C}_1 e^{2 u}\right)\right)}{8 \pi }~.
\end{align}
Let us now derive the corresponding eigenvectors. We start with the first eigenvalue $m_1^2$ as:
\begin{equation}
    M^2 \vec{\alpha}_1 = m_1^2 \vec{\alpha}_1 \Rightarrow M^2_{i1}\chi_1 + M^2_{i2}\chi_2 + M^2_{i3}\chi_3 = m^2_1 \chi_i~.\label{eigeneq}
\end{equation}
Due to the requirement that we would like to have a hierarchy in the masses and the internal geometry, we demand that $\langle\tau_{1}\rangle\ll \langle\tau_2, \tau_3, \mathcal{V}\rangle$. This hierarchy can be safely achieved by tuning the fluxes $\mathcal{W}_0$ and the $\mathcal{C}_i$ parameters, as these are given in the minima of equation \eqref{uv1}. Thus, we need to observe the scaling of the ratio between the entries of the mass matrix and the eigenvalue. 
\begin{align}
    &\dfrac{M^2_{11}}{m^2_1} \sim \mathcal{O}(1), \;  \dfrac{M^2_{12}}{m^2_1} \sim \mathcal{O}(\dfrac{\tau_1}{\tau_2}), \;  \dfrac{M^2_{13}}{m^2_1} \sim \mathcal{O}(\dfrac{\tau_1^2 \tau_2}{\mathcal{V}^2}),\notag\\
    &\dfrac{M^2_{12}}{m^2_1} \sim \mathcal{O}(\dfrac{\tau_2}{\tau_1}),\; \dfrac{M^2_{22}}{m^2_1} \sim \mathcal{O}(1), \;\dfrac{M^2_{23}}{m^2_1} \sim \mathcal{O}(\dfrac{\tau_1 \tau_2^2}{\mathcal{V}^2}),\notag\\
     &\dfrac{M^2_{31}}{m^2_1}\sim \mathcal{O}(\dfrac{\mathcal{V}^2}{\tau_1^2 \tau_2}),\; \dfrac{M^2_{32}}{m^2_1}\sim \mathcal{O}(\dfrac{\mathcal{V}^2}{\tau_1 \tau_2^2}),\;\dfrac{M^2_{33}}{m^2_1} \sim \mathcal{O}(1)~.\label{ratios}
\end{align}
The large hierarchy in the mixing is required for the thermalization procedure, since a parametrically small coupling cannot contribute sufficiently to maintain in thermal equilibrium the moduli with the MSSM thermal bath. Next, we are going to analyze the derivation of the first eigenvector, and the rest will follow applying the same recipe. For the heavy eigenvalue, the following constraints descend from equation \eqref{eigeneq}, taking into the ratios defined in \eqref{ratios}: 
\begin{align}
   & \dfrac{M_{11}^2}{m_1^2}\chi_1+\dfrac{M_{12}^2}{m_1^2}(\chi_2+\chi_3) = \chi_1\Rightarrow \chi_1 = \dfrac{M_{12}^2}{m_1^2-M_{11}^2}(\chi_1+\chi_3),\\
   & \dfrac{M_{21}^2}{m_1^2}\chi_1+\dfrac{M_{22}^2}{m_1^2}(\chi_2+\chi_3) = \chi_2 \Rightarrow \chi_3 =\chi_2 (-1+ \dfrac{m_1^2}{1+ \dfrac{M_{12}^2M_{21}^2}{M_{22}^2(m_1^2-M_{11}^2)}})~.
\end{align}
From these two equations and using the ratios in equation \eqref{ratios} , we can determine two components of the eigenvector, while the third will be specified by normalization of the eigenvectors. So:
\begin{align}
    \vec{\alpha}^T\cdot K \cdot \vec{\alpha}=\dfrac{1}{4\tau_1^2}\chi_1^2 + \dfrac{1}{4\tau_2^2}\chi_2^2 +&\dfrac{1}{4\tau_3^2}\chi_3^2 \cong \dfrac{1}{4\tau_2^2}(\chi_2^2+\chi_3^2)=1\Rightarrow (\chi_2^2+\chi_3^2) \cong 4 \tau_3^2\Rightarrow\\
    &\chi_2\cong \chi_3 \cong \sqrt{2}\tau_3,
\end{align}
where in the last step, we have applied our requirement that the vev of $\tau_1$ modulus is smaller compared to the other two moduli. This will be justified in the next section, where we explicitly derive the vev of the moduli at the stable vacuum. Consequently, the first eigenvector can be described by the following form:
\begin{align}
    \vec{\alpha}_1 = \{\chi_1,\; \chi_2,\; \chi_3\}~.
\end{align}
The other two eigenvectors characterize the degenerate light eigenvalues. So, we follow the same procedure to derive their scaling.
\begin{align}
   & \chi_1+\dfrac{M_{12}^2}{M_{11}^2}(\chi_2+\chi_3) = \dfrac{m_2^2}{M_{11}^2}\chi_1\Rightarrow \chi_1 = \dfrac{M_{12}^2}{m_2^2-M_{11}^2}(\chi_1+\chi_3),\\
   & \chi_1+\dfrac{M_{22}^2}{M_{11}^2}(\chi_2+\chi_3) =\dfrac{m_2^2}{M_{11}^2} \chi_2 \Rightarrow \chi_3 =\chi_2 (-1+ \dfrac{m_2^2 M_{21}^2(m_2^2-M_{11}^2)}{M_{22}^2(m_2^2-M_{11}^2)+ M_{12}^2M_{21}^2})~.
\end{align}
Similarly exploiting the ratios between the eigenvalues and the matrix elements at the vacuum, the components of the eigenvectors can be approximated by the following formula:
\begin{align}
    &\chi_3\cong \chi_2 \cong  \sqrt{2} \tau_3,\\
    \vec{\alpha}_2=\{\chi_1, \chi_3, \chi_2\}&,\; \vec{\alpha}_3=\{\chi_2, c\chi_3, \chi_1\},\; c\sim \mathcal{O}(1)~.
\end{align}
The constant $c$ is introduced in the computation of the "light" eigenvectors due to the degeneracy of the masses, and it can perceived as the splitting factor of the light eigenvalues. The diagonalization of the mass matrix and the ordering of the eigenvalues lead to the following pattern of mixing between the moduli $\tau_i$ and the normalized fields $\phi_i$. 
\begin{align}
    & \delta \tau_1 = \chi_{1}(\vec{a}_1) \delta \phi_1 + \chi_{1}(\vec{a}_2) \delta \phi_2 + \chi_{2}(\vec{a}_3) \delta \phi_3,\\
    & \delta \tau_2 = \chi_{2}(\vec{a}_1) \delta \phi_1 + c\chi_{3}(\vec{a}_2) \delta \phi_2 + \chi_{3}(\vec{a}_3) \delta \phi_3,\\
    & \delta \tau_3 = \chi_{3}(\vec{a}_1) \delta \phi_1 + \chi_{2}(\vec{a}_2) \delta \phi_2 + \chi_{1}(\vec{a}_3) \delta \phi_3~.
\end{align}
From the above mixing, we can clearly see a geometric separation of the internal geometry, which means that $\delta \tau_1 \sim \chi_2 (\vec{\alpha}_3) \delta \phi_3$ while $\delta \tau_{2,3} \sim \chi_3 (\vec{\alpha}_3) \delta \phi_{1,2}$. Regarding the ordering of the eigenvalues, they are given by this form $m^2_i = \{m^2_3,m_2^2,m_1^2 \}$. Turning our attention now on the coupling of the moduli to the gauge bosons $X$, these can be read from the gauge kinetic function \cite{hep-th/0610129}:
\begin{align}
    f_i = \dfrac{T_i^{MSSM}}{4\pi} + h_i(F)S,
\end{align}
where $T_i$ denotes the K{\"a}hler modulus on which the MSSM resides, while $h_i$ is a topological function of the internal world-volume fluxes of the compactification. Taking this into account, we can assume that the small modulus $\tau_1$ characterizes the visible sector, so the following term of the Lagrangian can be written as:
\begin{align}
    \mathcal{L} = -\dfrac{\tau_1}{M_{pl}}F_{\mu\nu}F^{\mu\nu} = -G_{\mu \nu}G^{\mu \nu} - \dfrac{\delta \tau_1}{\langle \tau_1\rangle M_{pl}}G_{\mu \nu}G^{\mu \nu}=-G_{\mu \nu}G^{\mu\nu}-\dfrac{\chi_2(\vec{\alpha}_3)}{ \langle \tau_1\rangle M_{pl}}\phi_3G_{\mu \nu}G^{\mu\nu},\quad \chi_2(\vec{\alpha}_3)= \sqrt{2}\tau_3,\label{coupling}
\end{align}
where in the last step, we expanded around the minimum of $\tau_1$ and normalized the gauge fields $G_{\mu\nu} = \sqrt{\langle \tau_1\rangle}F_{\mu \nu}$. Consequently, we can observe that the heaviest field will be thermalized under specific conditions with respect to the magnitude of the coupling, which will be addressed in subsequent sections. \par
During the expansion of the universe, a specific  particle species will remain in thermal equilibrium if the interaction rate $\Gamma$ with the particle content of the thermal bath is greater than the expansion rate $H\sim g_*^{1/2}T^2/M_{pl}$ of the universe, with $g_*$ being the degrees of freedom in the thermal bath. The processes that contribute to the above thermalization are $2 \leftrightarrow 2$ (scattering and annihilations) processes and $1 \leftrightarrow 2$ decays and inverse production. Starting with $2 \leftrightarrow 2$ processes, we can introduce the thermally averaged interaction rate
\begin{align}
    \Gamma \sim \dfrac{1}{\langle t_c \rangle}, \; t_c \sim \dfrac{1}{n \sigma v},
\end{align}
where $t_c$ is the mean time between collisions, n is the number density, $\sigma$ is the cross section and $v$ is the relative velocity between the particles of the thermal bath. The relativistic nature of the particles dictates the velocity to be of order $v\sim c\equiv 1$ and the number density $n\sim T^3$. We are interested in processes that include gravitational interactions, in order to see if under specific geometric enhancement through the moduli couplings, these interactions could become relevant for the thermalization of $2\leftrightarrow 2$ processes. The cross sections for the aforementioned interactions are given by \cite{Anguelova:2009ht}:
\begin{itemize}
    \item i) Diagram with two gravitational vertices:
    \begin{equation}
        \langle \sigma \rangle \sim \mu \dfrac{T^2}{M_{pl}^4} \Rightarrow \langle \Gamma \rangle \sim \mu \dfrac{T^5}{M_{pl}^4}~.
    \end{equation}
    \item ii) Diagram with one renormalizable and one gravitational vertex:
     \begin{equation}
        \langle \sigma \rangle \sim \sqrt{\mu} \dfrac{g^2}{M_{pl}^2} \Rightarrow \langle \Gamma \rangle \sim \sqrt{\mu} \dfrac{g^2T^3}{M_{pl}^2},
    \end{equation}
\end{itemize}
where $\mu$ is a moduli-dependent factor, more specifically the mixing of moduli with the gauge bosons, and $g$ is the gauge coupling of the theory. Comparing these interaction rates with the expansion rate of the universe, we can derive the freeze-out temperature of the particular particle species. The freeze-out temperatures for the cases described above are expressed as:
\begin{align}
    i)\;&\langle \Gamma \rangle > H \Rightarrow T > g_*^{1/6} \dfrac{M_{pl}}{\mu^{1/3}},\label{grvr}\\
    ii)\;&\langle \Gamma \rangle > H \Rightarrow T > g_*^{1/2} \dfrac{M_{pl}}{g^2\mu^{1/2}}~.\label{renver}
\end{align}
Turning our attention to decay processes, gravitational interactions mediated by massive particles at $T<M$ (with $M$ being the Planck mass) result in a decay rate of the following form:
\begin{equation}
    \Gamma \sim \mu \dfrac{m^3}{M_{pl}^2}~.
\end{equation}
In order to compare the above quantity with the expansion rate, we need to consider the thermally averaged decay rate:
\begin{equation}
    \langle \Gamma^{\prime} \rangle = \Gamma \dfrac{m}{\langle E \rangle},
\end{equation}
where we introduced a relativistic factor, in order to switch from the rest frame to the laboratory frame. This factor in the relativistic regime is given by $T\gtrsim m$, $\gamma = \langle E\rangle /m =\langle T \rangle/m$, while in the opposite limt for $T \lesssim m$, $\gamma \sim 1~.$ Regarding the inverse decay, they require temperatures higher than the mass $T \gtrsim m$, otherwise the final states do not have enough energy to reverse the decay. To summarize the requirements for the current case study, we write down the decay rates at $T \lesssim M$:
\begin{align}
    &i)\; \langle \Gamma_{1\rightarrow 2} \rangle \cong  \begin{cases}
    \mu \dfrac{m^4}{M_{pl}^2T},\; T \gtrsim m,\\
    \mu \dfrac{m^3}{M_{pl}^2},\; T \lesssim m~.
    \end{cases}\\
    &ii)\; \langle \Gamma_{2\rightarrow 1}\rangle \cong
    \begin{cases}
        \mu \dfrac{m^4}{M_{pl}^2T},\; T \gtrsim m,\\
        \mu\dfrac{m^3}{M_{pl}^2}(\dfrac{m}{T})^{3/2}e^{-m/T},\; T\lesssim m,
    \end{cases}
\end{align}
where, again, the $\mu$ factor is modulus dependent and the Boltzmann suppression $e^{-m/T}$ is introduced for processes that have insufficient energy to recreate the particle ($T\lesssim m$). Therefore, the thermal equilibrium for $T \gtrsim m$ is maintained only if:
\begin{align}
    \langle \Gamma\rangle >H \Rightarrow 1< \dfrac{T}{m}< (\dfrac{\mu\; m}{g_*^{1/2} M_{pl}})^{1/3}~.\label{invdec}
\end{align}

\section{Achieving a dS space}

On the basis of the preceding analysis, we observe that the previous vacuum is AdS and it is necessary to introduce an uplifting D-term \cite{Burgess:2003ic, Haack:2006cy, Achucarro:2006zf}. The relevant term we could implement as an uplifting term scales as:
\begin{equation}
    V_{up} = \dfrac{d}{\mathcal{V}^2}~.
\end{equation}
The effective potential, under consideration, can now be formulated as:
\begin{align}
     V_{eff} =&\; \frac{3 \mathcal{W}_0^2 g_s (-16 \eta +\xi +4 \eta  \log (\mathcal{V}))}{4 \mathcal{V}^3} -\frac{\mathcal{W}_0^2 g_s \left(\frac{\mathcal{C}_3 \mathcal{V}^3}{\tau _1 \tau _2}+\mathcal{C}_1 \tau _1\mathcal{V}+\mathcal{C}_2 \tau _2 \mathcal{V}\right)}{8 \pi  \mathcal{V}^5} - \notag \\
     & -\frac{3 \lambda  \mathcal{W}_0^4 \sqrt{g_s} \left(\tau _1 \tau _2+\frac{\mathcal{V}^2}{\tau_1}+\frac{\mathcal{V}^2}{\tau _2}\right)}{8 \pi ^2 \mathcal{V}^5} +\dfrac{d}{\mathcal{V}^2}~.
\end{align}
This potential can effectively stabilize all the moduli at Minkowski/dS spaces without introducing any tachyonic directions. Again we can change our basis to find the vacua in the orthonormal  basis, where the potential is written as:
\begin{align}
    V_{eff} =&\; \frac{3}{4} e^{-3 \sqrt{\frac{3}{2}} t} \mathcal{W}_0^2 g_s \left(\xi +2 \eta  \left(\sqrt{6}t-8\right)\right) -\frac{\mathcal{W}_0^2 g_s e^{-5 \sqrt{\frac{2}{3}} t-u-\frac{2 v}{\sqrt{3}}} \left(\mathcal{C}_3e^u+e^{\sqrt{3} v} \left(\mathcal{C}_2+\mathcal{C}_1 e^{2u}\right)\right)}{8 \pi } - \notag \\
    & -\frac{3 \lambda  \mathcal{W}_0^4 \sqrt{g_s}\left(e^{u+\sqrt{3} v}+e^{2 u}+1\right) e^{-\frac{11t}{\sqrt{6}}-u-\frac{v}{\sqrt{3}}}}{8 \pi ^2} + d e^{-\sqrt{6} t}~.\label{VT0}
\end{align}
The first modulus $u$ can be stabilized at:
\begin{align}
    \dfrac{\partial V_{eff}}{\partial u} = 0 \Rightarrow u_{min} = \dfrac{1}{2} \;\log( \frac{\pi  \mathcal{C}_2 \sqrt{g_s} e^{\frac{t}{\sqrt{6}}+\frac{2 v}{\sqrt{3}}}+3 \lambda 
   \mathcal{W}_0^2}{\pi  \mathcal{C}_1 \sqrt{g_s} e^{\frac{t}{\sqrt{6}}+\frac{2 v}{\sqrt{3}}}+3 \lambda 
   \mathcal{W}_0^2} )~.
\end{align}
Regarding the volume direction $t$, it can be effectively described by leading order corrections (log-loop corrections and D-terms) as opposed to our previous analysis. In this procedure, we assumed that the minima of $u,v$ are mainly given by \eqref{uv1}, since the subleading $F^4$ corrections are a minimal effect. This leads us to the following formula:
\begin{equation}
     \dfrac{\partial V_{eff}}{\partial t} = 0 \Rightarrow t_{min} \cong \;\frac{-12 \sqrt{6} \eta\;  W_{0/-1}\left(\frac{2 d e^{\frac{13}{3}-\frac{\xi }{4 \eta }}}{9 \eta  \mathcal{W}_0^2g_s}\right)+52 \sqrt{6} \eta -3 \sqrt{6} \xi }{36 \eta },
\end{equation}
where $W_{0/-1}$ stands for the Lambert function, where the upper branch denotes the maximum of the potential and the lower branch the minimum. As for the last direction, it can be approximated by:
\begin{equation}
    \dfrac{\partial V_{eff}}{\partial v} = 0 \Rightarrow v_{min} \cong -\dfrac{\sqrt{2}t}{2} - \frac{\log \left(\frac{1024 \pi ^6 \mathcal{C}_1^5 \mathcal{C}_2^5 \mathcal{C}_3^2 g_s^3}{\left(\left(\mathcal{C}_1+\mathcal{C}_2\right) \lambda  \mathcal{W}_0^2+3 \pi (-\mathcal{C}_1 \mathcal{C}_2 \mathcal{C}_3)^{2/3} e^{\frac{t}{\sqrt{6}}}
   \sqrt{g_s}\right){}^6}\right)}{2\sqrt{3}}~.
\end{equation}
It is evident that the volume is mainly expressed by the log-loop effects, while the subleading corrections (winding loop effects) stabilize the transverse directions, generating in that way a geometrical hierarchy both in the masses and mixing. While our analysis for the mixing in the previous section performed taking into account only the F-term, the introduction of the D-term affects only the eigenvalue of the lightest mode. Figure 1. and Figure 2. depict graphically the global vacuum along the orthonormal basis $(t,u,v)$, where Table 1. and Table 2. provide the parameters used for the current vacuum along with the vevs and masses of the moduli.
\begin{figure}[H]
\centering
\begin{subfigure}{}
  \centering
  \includegraphics[scale=0.65]{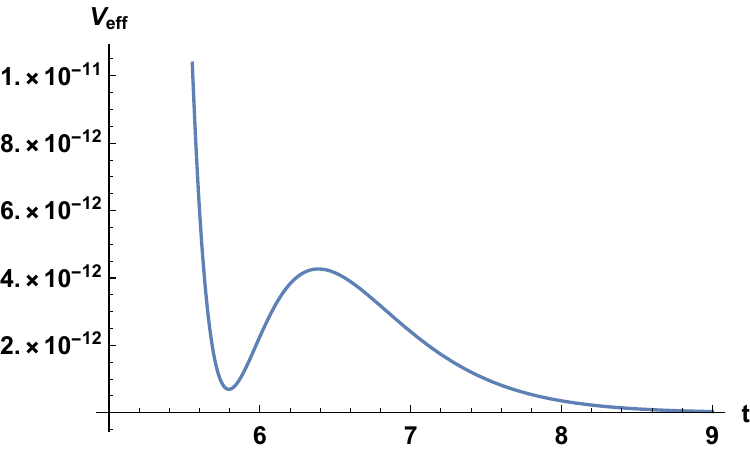}
  \label{fig:sub1}
\end{subfigure}%
\begin{subfigure}{}
  \centering
  \includegraphics[scale=0.65]{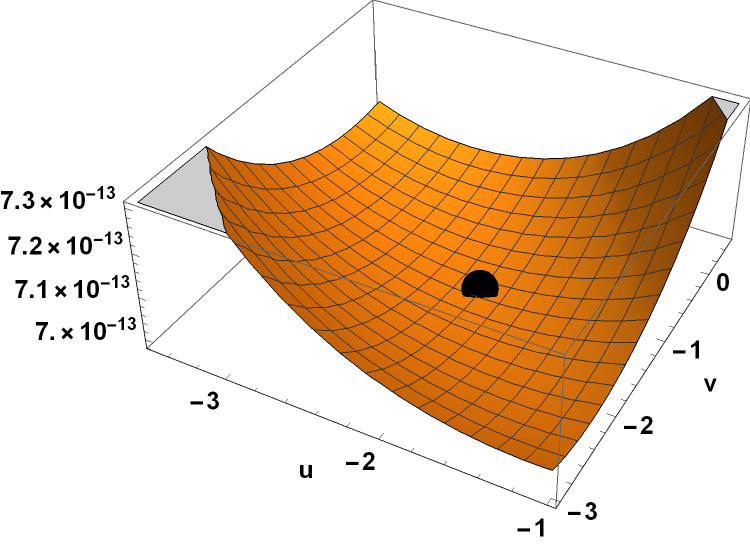}
  \label{fig:sub2}
\end{subfigure}
\caption{Vacuum of the potential \eqref{VT0} along each direction. The black dot represents the minimum.}
\end{figure}
\begin{table}[H]
\begin{center}
\begin{tabular}{|c|c|c|c|c|c|c|c|}
\hline
  \cellcolor[gray]{0.9}$g_s$ & \cellcolor[gray]{0.9}$|\mathcal{W}_0|$ & \cellcolor[gray]{0.9}$\hat{\xi}$  & \cellcolor[gray]{0.9}$|\hat{\eta}|$  & \cellcolor[gray]{0.9}$|\lambda|$  & \cellcolor[gray]{0.9}$d$ & \cellcolor[gray]{0.9}$|\mathcal{C}_1|$ & \cellcolor[gray]{0.9}$|\mathcal{C}_2|=|\mathcal{C}_3|$\\
  \hline
  0.1 & 1 & 5 & 0.6 & $1\times 10^{-3}$ &$1.52 \times 10^{-3}$ & 0.5 & $1\times 10^{-3}$\\
\hline\hline 
\end{tabular}
\end{center}
\caption{The parameters used in the plots above.}
\end{table}%
\begin{table}[H]
\begin{center}
\begin{tabular}{|c|c|c|c|c|c|c|c|}
\hline
  \cellcolor[gray]{0.9}$\tau_1$ & \cellcolor[gray]{0.9}$\tau_2$ & \cellcolor[gray]{0.9}$\mathcal{V}$  & \cellcolor[gray]{0.9}$m^2_{\tau_1}$  & \cellcolor[gray]{0.9}$m^2_{\tau_2}$  & \cellcolor[gray]{0.9}$m^2_{\mathcal{V}}$\\
  \hline
  3 & 650 & 1208 & $1.5\times 10^{-10}$ & $7.7\times 10^{-14}$ &$2.4\times 10^{-15}$ \\
\hline\hline 
\end{tabular}
\end{center}
\caption{Vevs of the moduli at the vacuum and their corresponding masses.}
\end{table}%
\noindent It is important for someone to observe that the D-terms only shift the eigenvalues at smaller values and never create a tachyonic direction. Regarding the observed asymmetry of the $\mathcal{C}_i$ parameters, this aspect favors the small vev of the corresponding modulus while it enhances the thermalization effects of this direction as it will be proved in the following section. Previous studies have assumed identical coefficients, although an explicit computation is needed to unravel their origin and effects after the stabilization of complex-structure moduli. Finally, the validity of the existence of a dS vacuum is translated as an upper bound of the uplift parameter $d$. This upper bound can be found from the Lambert function and the requirement that both branches exist ($W_{0/-1}(x),\;-1/e\leq x< 0$).
\begin{equation}
    -\dfrac{1}{e} \leq \dfrac{2d e^{\frac{13}{3}-\frac{\xi}{4\eta}}}{9g_s \mathcal{W}_0^2 \eta}<0 \Rightarrow -\dfrac{9}{2} g_s \mathcal{W}_0^2 \eta e^{-\frac{16}{3}+\frac{\xi}{4\eta}}\geq d > 0~.
\end{equation}
As for the lower bound, this can be derived by demanding the positivity of the potential at the global minimum. The potential there can approximated by:
\begin{equation}
    V_{eff}^{min} \cong \dfrac{2d^3 (2+3  W_{0}\left(\frac{2 d e^{\frac{14}{3}-\frac{\xi }{4 \eta }}}{27 \eta \mathcal{W}_0^2 g_s}\right))}{729 g_s^2 \mathcal{W}_0^4\eta^2  W_{0}\left(\frac{2 d e^{\frac{14}{3}-\frac{\xi }{4 \eta }}}{27 \eta \mathcal{W}_0^2 g_s}\right)^3} + \mathcal{O}(\mathcal{C}_i,\lambda)~.
\end{equation}
Now, the lower bound is given by demanding that the parentheses in the numerator has negative sign, in order to compensate the negative sign from the Lambert function in the denominator:
\begin{equation}
    \frac{2 d e^{\frac{14}{3}-\frac{\xi }{4 \eta }}}{27 \eta \mathcal{W}_0^2 g_s}< -\dfrac{2}{3}e^{-\frac{2}{3}}\Rightarrow d \gtrsim -3g_s\mathcal{W}_0^2 \eta \; e^{-\frac{15}{3}+\frac{\xi}{4\eta}}~.
\end{equation}
Consequently, the final bounds on the can expressed by the following very stringent inequality:
\begin{equation}
    -\dfrac{9}{2} g_s \mathcal{W}_0^2 \eta \; e^{-\frac{16}{3}+\frac{\xi}{4\eta}}\geq d \gtrsim -3g_s\mathcal{W}_0^2 \eta \; e^{-\frac{15}{3}+\frac{\xi}{4\eta}}~.\label{dSbound}
\end{equation}

\section{Finite temperature induced vacua and related cosmological implications}
Moving on, the next thing is to consider if the heaviest modulus can thermalize at sub-planckian temperatures. The $\tau_1$ modulus describes the heaviest normalized field $\phi_3$ as indicated by the analysis of Section 2., which means that finite temperature corrections could potentially be introduced along this direction at the level of the effective potential. Before studying their dynamics, we need to establish the point in cosmological times, in which the thermal equilibrium is achieved. Scattering or pair production processes, either with two gravitational vertices (equation \eqref{grvr}) or with one renormalizable and one gravitional vertex (equation \eqref{renver}), can achieve thermal equilibrium of particles species with a thermal bath above the following freeze-out temperatures $T_f$:
\begin{align}
    T^{grav} > T_f \equiv g_*^{1/6} \dfrac{M_{pl}}{2^{2/3}\tau_3^{4/3}}, \quad T^{renor} > T_f \equiv g_*^{1/2} \dfrac{M_{pl}}{2g^2 \tau_3}~,
\end{align}
where for the modulus dependent factor $\mu$, we used the coupling of $\phi_3$ field to the gauge bosons, as given by equation \eqref{coupling} $g_{\phi_3 XX}\sim \chi_2(\vec{\alpha}_3) = \sqrt{2}\tau_3$, since the vev of $\langle\tau_1\rangle$ modulus is of order one. Both of these temperatures are below the Planck scale and in principle thermalize the vacuum after reheating $T_{rh}$. This implies an indirect lower bound on reheating, since a correct implementation of the inflationary scenario would require $T_f <T_{rh}$, otherwise finite temperature corrections have no effect. A similar argument can be applied for the decay and inverse decay processes, where we demand that the temperature $T$ exceed the mass of $\phi_3$, in order to probe $X + X\rightarrow \phi_3$ processes:
\begin{equation}
    T>m_{\phi_3} \equiv \big(12 d\; e^{ 2 W_0\left(\frac{2 d e^{\frac{13}{3}-\frac{\xi }{4 \eta }}}{9 \eta \mathcal{W}_0^2g_s}\right)+\frac{\xi }{2 \eta }-\frac{26}{3}} -\frac{4 d^3 \left(5 W_0\left(\frac{2 d e^{\frac{13}{3}-\frac{\xi }{4 \eta }}}{9 \eta  \mathcal{W}_0^2
   g_s}\right)+1\right)}{27 \eta ^2 \mathcal{W}_0^4 g_s^2 W_0\left(\frac{2 d e^{\frac{13}{3}-\frac{\xi }{4
   \eta }}}{9 \eta  \mathcal{W}_0^2 g_s}\right){}^3})^{1/2}~.\label{activation}
\end{equation}
Now, this temperature has to be confronted by the equilibrium temperature for inverse decays, whose definition descends from equation \eqref{invdec}. Since the $\mu= \chi_2(\vec{\alpha}_3)^2$ factor denotes the coupling for the field with the gauge bosons in this process, we can write:
\begin{equation}
    T<T_{eq} \equiv (2^{1/3}e^{\frac{2^{1/3} \left(\sqrt{2} t-2 v\right)}{3 \sqrt{3}}} \dfrac{2^{1/3}m_{\phi_3}^{1/3}}{M_{pl}^{1/3}g_*^{1/6}})m_{\phi_3} \cong (\underbrace{\frac{2^{1/3} e^{ \frac{5}{27} \left(-12 W_0\left(\frac{2 d e^{\frac{13}{3}-\frac{\xi }{4 \eta }}}{9 \eta \mathcal{W}_0^2 g_s}\right)-\frac{3 \xi }{\eta }+52\right)}}{g_*^{1/6}}\dfrac{m_{\phi_3}^{1/3}}{{M_{pl}^{1/3}}}}_{\text{y}}) m_{\phi_3} ,
\end{equation}
which is clearly greater than the activation temperature \eqref{activation} for the decays even at leading order, due to the fact that $y>1$ at the dS vacuum presented in Table 2. Inevitably, decay and inverse decays do contribute to the thermalization of the modulus. \par
Regarding the finite temperature corrections at the effective potential, it was pointed out in \cite{Anguelova:2009ht} that 1-loop effects are subleading compared to 2-loop corrections taking into account that the MSSM is supported by one modulus. This also holds in our model, since the mass scale of both fermions and bosons does not bring any new moduli dependence in the effective theory. This could be understood from the fact that these eigenvalues are derived from the mass matrix of the current model. We schematically prove our argument by writing the potential up to 1-loop effects as:
\begin{align}
V_{eff} = V(T=0) + \dfrac{T^2}{M_{pl}^2}( m^2_{\phi_3})\sim  e^{-3 \sqrt{\frac{3}{2}} t}\mathcal{W}_0^2 (1 + \underbrace{\dfrac{T^2}{M_{pl}^2}}_{\ll 1}) + ...,\;   
\end{align}
where in the above equation we used the $t$ dependent part of the T=0 potential at \eqref{VT0}, while for the mass of the modulus we applied equation \eqref{massphi}. Thus, in order to bring into the table corrections that could compete with the potential at zero temperature, we examine the 2-loop thermal corrections. Since $\tau_1$ modulus is thermalized, from now on we consider the following potential: 
\begin{equation}
    V_{tot} = V_{eff} + V_{up} +T^4 \left(\kappa_1 g_{MSSM}^2+\kappa_2 g_{\phi_3 XX}^2 m_{\phi_3 }^2\right) \cong  V_{eff} + V_{up} +T^4( \dfrac{1}{\tau_1} + (\chi_2(\vec{\alpha}_3))^2 Tr[M^2])~,\label{Vtot}
\end{equation}
where we assume that $\kappa_{1,2}$ are constants of order one and are absorbed. This can be justified by considering the fact that the MSSM are created by $D_7$ branes associated with the $\tau_1$ modulus. Then, the coefficient $\kappa_1$ is defined by the rank of the gauge group $SU(N)$:
\begin{equation}
    \kappa_1 = \dfrac{(N^2-1)(N-N_f)}{64},
\end{equation}
where $N_f$ is the number of fermions \cite{Kapusta:2007xjq}. Despite the fact that previous studies have not managed to find new vacua due to thermal effects, in the current model we are going to show that new thermally-induced vacua can be obtained. In our approach, it will be shown that the T-dependent contributions are comparable to the dominant contribution of the effective potential at $T=0$. This contradicts previous studies\cite{Anguelova:2009ht}, where the non-perturbative corrections along the thermalized direction are parametrically larger than the finite temperature effects. On the contrary, in our perturbative stabilization scheme, the transverse directions to the volume are stabilized by subleading effects (winding loop corrections), which can compete with the finite-T induced corrections, leading to a reshape of the potential. Thus, we are obliged to take into account the full re-stabilization of the total potential and search for vacua that are connected to the $T=0$ dS vacuum. This argument will become clear, when we will prove that the compactification volume only slightly shifts while the displacement of the other moduli is more robust. The new ingredients could in principle provide us with further bounds on the free parameters of the model, while it could be possible to associate the decompactification temperature $T_{max}$ with the uplifted T-dS vacua. The shift in the vevs of the moduli can be achieved by re-stabilizing the potential. Starting from the $u$ direction:
\begin{align}
    \dfrac{\partial V_{tot}}{\partial u} \cong 0 \Rightarrow u_{min} \cong \dfrac{1}{2} \;\log \left(\frac{2 \pi  \mathcal{C}_2 \mathcal{W}_0^2 g_s e^{\frac{t}{\sqrt{6}}+\frac{2 v}{\sqrt{3}}}+3
   \lambda  \mathcal{W}_0^4 \sqrt{g_s}-16 \pi ^2 e^{3 \sqrt{\frac{3}{2}} t} T^4}{2 \pi  \mathcal{C}_1\mathcal{W}_0^2 g_s e^{\frac{t}{\sqrt{6}}+\frac{2 v}{\sqrt{3}}}+3 \lambda  \mathcal{W}_0^4
   \sqrt{g_s}}\right)~.
\end{align}
We can see that a new factor proportional to the temperature is appended in the previously derived vacuum. This factor comes with an exponent $\sim T^4 e^{-3\sqrt{3}t/2}$ greater than the leading order exponent $\sim \mathcal{C}_2 e^{t/\sqrt{6}}$ of the previous vacuum solution. This indicates a large shift due to finite temperature corrections. Moving to the volume direction $t$, an approximate formula at leading order is given by:
\begin{align}
    \dfrac{\partial V_{tot}}{\partial t} = 0 &\Rightarrow \frac{e^{-3 \sqrt{\frac{3}{2}} t} \left(-24 d e^{\sqrt{\frac{3}{2}} t} \left(8 T^4 e^{\frac{2 \left(\sqrt{2} t-2 v\right)}{\sqrt{3}}}+1\right)-9 \mathcal{W}_0^2 g_s \left(-52 \eta +3 \xi +6 \sqrt{6} \eta t\right)-8 T^4 e^{\frac{7 t}{\sqrt{6}}-u-\frac{v}{\sqrt{3}}}\right)}{4 \sqrt{6}} + \notag \\
    & + \frac{5 \mathcal{W}_0^2 g_s e^{-5 \sqrt{\frac{2}{3}} t-u-\frac{2 v}{\sqrt{3}}} \left(\mathcal{C}_3 e^u+e^{\sqrt{3} v} \left(\mathcal{C}_2+\mathcal{C}_1 e^{2 u}\right)\right)}{4 \sqrt{6} \pi } \cong 0~.
\end{align}
\begin{align}
    t_{min} \cong \frac{-12 \sqrt{6} \eta  W_{0}\left(-\frac{e^{\frac{13}{3}-\frac{\xi }{4 \eta }} \left(\sqrt{\frac{2}{\pi }}\mathcal{W}_0 \sqrt{-\mathcal{C}_1 g_s T^4 }-2 d\right)}{9 \eta  \mathcal{W}_0^2 g_s}\right)+52 \sqrt{6} \eta -3 \sqrt{6} \xi }{36 \eta }~.\label{tmin}
\end{align}
It is evident that the thermally induced vacuum slightly shifts the volume due to the suppressed term of order $\sim T^2$. In an interesting way at high temperatures, the winding loop effects are the mediator of the novel contributions along the visible sector, a fact that is analogous to the role of loop-induced threshold corrections in heterotic string models \cite{hep-th/9502077, Atick:1988si, Catelin-Jullien:2007ewh}. Moreover, the $\mathcal{C}_1$ parameter, that accompanies $T$, is a loop effect along the $\tau_1$ direction and it brings out the necessary asymmetry needed along the winding loop effects that was mentioned before in the stabilization procedure. Despite being a fair assumption in our approach, the emergence of this asymmetry can be traced back to the dynamics of the complex structure moduli and the geometric properties of the background geometry. This behavior shares some interesting features with the work of \cite{ Liu:2011nw}, where the scalars are stabilized dynamically during the thermal cosmological evolution of the universe. Regarding the $v$ direction, we apply the approximate expressions for the minima of transverse directions and due to the complexity of the equations, numerically we find that:
\begin{equation}
    \dfrac{\partial V_{tot}}{\partial v}\cong 0 \Rightarrow v_{min} \cong - 4 v_{min}(T=0)~.
\end{equation}
This thermal displacement along the $u,v$ directions, which in practice parametrize the dynamics of $\tau_1,\tau_2$ moduli, can be traced back to the coupling of the MSSM branes $g^2_{MSSM} \sim \tau_1^{-1}\sim e^{-u-v/\sqrt{3}}$. This coupling exchanges the behavior of the vev between the $\tau_1$ and $\tau_2$ directions,  since the thermal effects are turned on along the "small" modulus $\tau_1$ and the compactification volume is fixed by the leading order effects. Thus, the above behavior for the stability of the vacuum is expected. To support our argument, the following plots (Figure 2. and Figure 3.) illustrate the vacua along each direction, to validate our approximations.\par
\begin{figure}[H]
    \centering
    \includegraphics[scale=0.7]{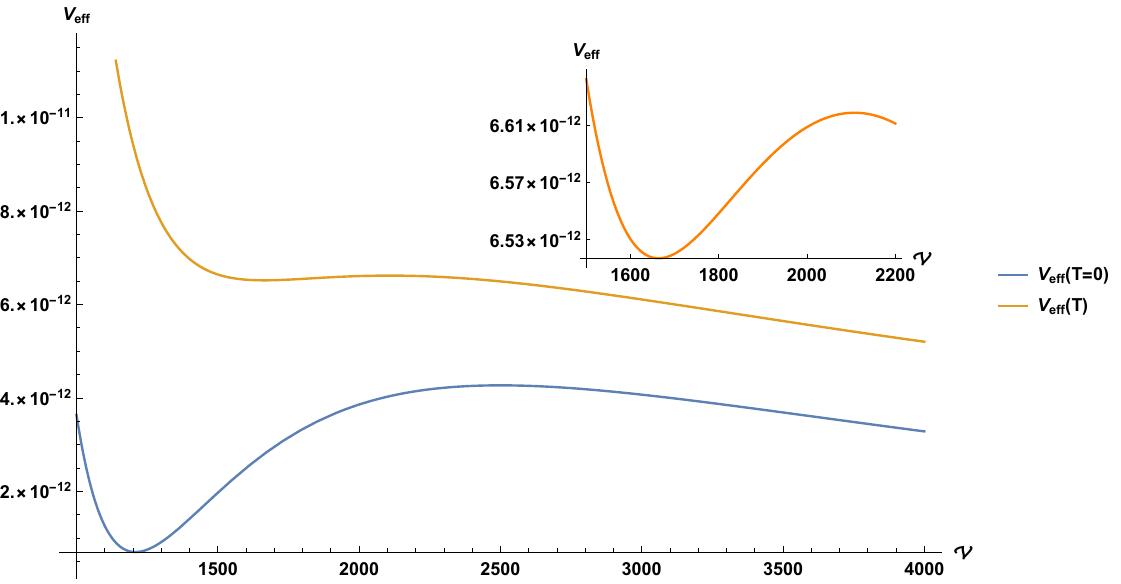}
    \caption{Plot of the potential \eqref{Vtot} along the volume direction under thermal effects.}
\end{figure}
\begin{figure}[H]
    \centering
    \includegraphics[scale=0.6]{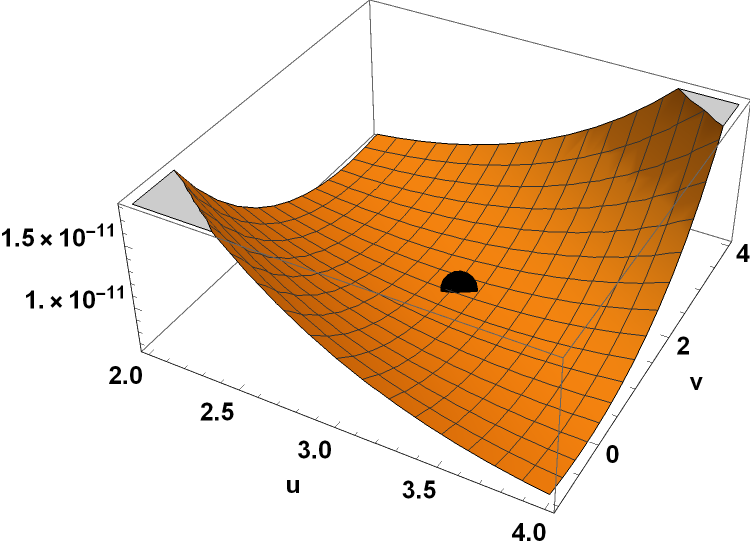}
    \caption{Plot of the potential \eqref{Vtot} along the transverse space under thermal effects.}
\end{figure}
\begin{table}[H]
\begin{center}
\begin{tabular}{|c|c|c|c|c|c|c|c|}
\hline
  \cellcolor[gray]{0.9}$\tau_1$ & \cellcolor[gray]{0.9}$\tau_2$ & \cellcolor[gray]{0.9}$\mathcal{V}$  & \cellcolor[gray]{0.9}$m^2_{\tau_1}$  & \cellcolor[gray]{0.9}$m^2_{\tau_2}$  & \cellcolor[gray]{0.9}$m^2_{\mathcal{V}}$ & \cellcolor[gray]{0.9}$T$ & \cellcolor[gray]{0.9}$T_{max}$ (GeV)\\
  \hline
  7100 & 17 & 1656 & $7\times 10^{-12}$ & $4.5\times 10^{-15}$ &$2 \times 10^{-21}$ & 0.01075 & $3 \times 10^{16}$\\
\hline\hline 
\end{tabular}
\end{center}
\caption{Vevs of the moduli at the vacuum and their corresponding masses. $T_{max}$ is the decompactification temperature.}
\end{table}%
\noindent As is already known in the literature, finite temperature corrections can destabilize the compactification above a critical point. Therefore, it is important to characterize the decompactification temperature $T_{max}$, which differs from the rough estimate that states that the critical temperature is given by $T_{max} \sim V_b^{1/4}$ with $V_b$ denoting the height of the potential barrier at $T=0$. Now, the maximum temperature can be extracted from the maximum point in the barrier which is given by the $W_{-1}$ branch of the Lambert function:
\begin{equation}
    t_{max} \cong \frac{-12 \sqrt{6} \eta  W_{-1}\left(-\frac{e^{\frac{13}{3}-\frac{\xi }{4 \eta }} \left(\sqrt{\frac{1}{\pi }}\mathcal{W}_0 \sqrt{-\mathcal{C}_1 g_s T^4 }-2 d\right)}{9 \eta  \mathcal{W}_0^2 g_s}\right)+52 \sqrt{6} \eta -3 \sqrt{6} \xi }{36 \eta }~.
\end{equation}
A few comments need to be made here regarding the nature of this solution. Firstly, the maximum is expressed by the lower branch of the Lambert function, while the prefactor of temperature effects slightly decreases compared to the minimum vev \eqref{tmin}. This fact can be understood as a hint that the thermal effects push the solutions close to the boundary of the range of validity of the Lambert function, which physically means that the temperature has the tendency to flatten the potential. By the above definition, we can extract the decompactification temperature $T_{max}$ in terms of the free parameters, through requiring a real solution with the argument of the Lambert function satisfying $x \geq -1/e $:
\begin{equation}
    T^2_{max} \leq \frac{ \sqrt{\pi }}{\sqrt{-\mathcal{C}_1 g_s}}(\frac{2 d}{\mathcal{W}_0 }+9 \eta \; g_s\mathcal{W}_0\; e^{\frac{\xi }{4 \eta }-\frac{16}{3}}) ~.
\end{equation}
This bound clearly implies that the decompactification temperature is well below the Planck scale. It is worth noticing also that the number of fluxes $\mathcal{W}_0$ do affect severely the cosmological evolution, since exponentially small fluxes would lead to a smaller decompactification temperature. This can be inferred from equation \eqref{dSbound}, where a smaller number of fluxes require smaller uplift $d$ and this consequently would lead to decompactification for high scale inflation. Thus, the above bound could be recognized, also, as an indirect cue to unravel the correlation between fluxes and inflation scale in type IIB LVS compactifications. One last check, to validate and quantify our claim, is to check the "thermal" force $R$ that drives the displacement of the moduli, and it is given by the steepness of the potential at the thermal vacuum:
\begin{align}
    R = \left| \dfrac{\partial_{\tau_1}V_T}{\partial_{\tau_1}V_0}\right| \sim 1~.\label{Tforce}
\end{align}
In the above equation we have used equation \eqref{Vtot}, where $V_0$ the $T=0$ part for the potential while $V_T$ describes the finite temperature corrections. Since the thermal energy can compete with the zero temperature potential, the re-stabilization of the effective theory was required and it offers an alternative to the cosmological evolution of the universe. Previous analyses of finite temperature effects in the LVS assumed that the modulus which couples to the thermal bath is stabilized by non-perturbative contributions to the superpotential. In this case, the stabilizing force dominates over thermal corrections, enforcing the hierarchy
\begin{equation}
\partial V_{T} \ll \partial V_{\mathrm{0}} \, .
\end{equation}
In perturbative LVS models, where K\"ahler moduli are stabilized by loop--induced and higher--derivative corrections rather than non--perturbative effects, this hierarchy could in principle fail. As a consequence, thermal corrections can compete with the stabilizing potential, leading to the appearance of thermally induced vacua. This regime lies outside the assumptions of previous analyses and has not been explored before. As a result, the standard conclusion that finite--temperature effects cannot significantly reshape the vacuum structure does not apply in perturbative LVS compactifications.\par
Finally, let us comment on the decays of the modulus $\phi_3$ during radiation domination. Since the volume direction provides a more fertile ground to study inflation due to its shape, we assume that $\phi_3$ is not responsible for initial reheating after the inflationary era. These decays could, in principle, wash out any previous cosmological relics if the modulus dominated era $T_{dom}$ precedes the decay temperature $T_D$ of the modulus. It is imperative to explore this scenario, because a modulus dominated era has important implications on the cosmological history concerning the baryon asymmetry, dark radiation \cite{Baer:2022fou,  Basiouris:2024buq, Cicoli:2022uqa} and dark matter abundances \cite{Ling:2025nlw, WileyDeal:2023trg}. The correct implementation of an inflationary scenario in the current framework requires that $T_f<T_{rh}<T_{max}$, while a correct estimate/computation of baryon asymmetry and Big-Bang nucleosynthesis is beyond the scope of this work. To this point, we need to compute the freeze-out temperature for the $2\rightarrow2$ processes, which can be taken out of equations \eqref{grvr}, \eqref{renver} where the $\mu$ factor is $\mu = (\chi_2(\vec{\alpha}_3))^4=8\tau_3^4$ \footnote{Each vertex $\phi F_{\mu \nu}F^{\mu\nu}$ carries two derivatives (two powers of momentum) and for a process $2\leftrightarrow2$ its amplitude $|A|^2$ scales as $|A|^2\sim g^4$. Here we use the reduced Plankc mass $M_{pl} = 2.4\times 10^{18}\; \text{GeV}~.$}:
\begin{equation}
    T_f > \frac{M_{pl} \left(\tau _1 \tau _2\right){}^{4/3} g_*^{1/6}}{2^{2/3} \mathcal{V}^{8/3}} \gtrsim  10^{15} \;\; \text{GeV}~,
\end{equation}
where the parameters used are listed in Table 2. and $g_*\sim 100$. Decays and inverse decay processes provide a similar bound on the order of magnitude for their freeze-out temperatures. Clearly, this bound indicates that finite temperature corrections for moderate values of volume tend to favor a high scale of decompactification temperature, as it is argued also in \cite{Anguelova:2009ht}. In addition, we prove in the final part of this section that decays of thermalized moduli cannot dilute the pre-existing relic abundancies in the current setup. Firstly, we compute the temperature at which the modulus starts to oscillate:
\begin{equation}
    T_{osc} \sim g_*^{-1/4} \sqrt{m_{\phi_3} M_{pl}}~.
\end{equation}
The energy density of the oscillating modulus can be characterized by its mass and initial amplitude of oscillation, and it can be written as:
\begin{equation}
    \rho_{\phi_3} \sim m^{2}_{\phi_3} \langle \tau_1\rangle^2~.
\end{equation}
Now, we can compare it with the energy density of radiation at the oscillating temperature, which scales as $\sim T^4$:
\begin{equation}
    (\dfrac{\rho_{\phi_{3}}}{\rho_{r}})|_{T_{osc}} \sim \dfrac{m^2_{\phi_3} \langle \tau_1\rangle^2}{g_* T_{osc}^4}\sim \dfrac{\langle \tau_1\rangle^2}{M_p^2}~.
\end{equation}
At the moment where the two energy densities are comparable, this signals the modulus domination temperature $T_{dom}$:
\begin{equation}
    (\dfrac{\rho_{\phi_{3}}}{\rho_{r}})|_{T_{dom}} \sim 1~.
\end{equation}
Using all the above, we can derive the domination temperature and define it as:
\begin{equation}
    T_{dom}(\dfrac{\rho_{\phi_{3}}}{\rho_{r}})|_{T_{dom}} \sim T_{osc}(\dfrac{\rho_{\phi_{3}}}{\rho_{r}})|_{T_{osc}}\Rightarrow T_{dom} \sim T_{osc} \dfrac{\langle \tau_1\rangle^2}{M_{pl}^2}~.
\end{equation}
Consequently, the following ratio can determine if the cosmological evolution includes a modulus dominated cosmological era:
\begin{equation}
    \dfrac{T_D}{T_{dom}} \sim \dfrac{g_{*}^{-1/4}\sqrt{\frac{g^2m_{\phi_3}^3}{64\pi M_{pl}}}}{g_*^{-1/4} \dfrac{\langle \tau_1\rangle^2}{M_{pl}^2}\sqrt{m_{\phi_3} M_{pl}}} \sim \dfrac{M_{pl}m_{\phi_3}}{\langle \tau_1 \rangle^2}\gg 1~.
\end{equation}
Obviously, the thermalized modulus decays well before its energy density starts to dominate the universe. The cosmological implications of such a result are directly related to early universe phenomena, such as baryon asymmetry and Big-Bang nucleosynthesis. Our result is in sharp contrast with moduli–dominated cosmologies, where late decays typically erase any previously generated asymmetry and require new mechanisms for baryogenesis \cite{Allahverdi:2010im, Dutta:2010sg, Allahverdi:2010rh, Higaki:2012ba, Garcia:2013bha}. Thus, given our high scale freeze out temperature and the need for a high scale inflation, various baryogenesis or leptogenesis can be implemented incorporating heavy Majorana neutrinos \cite{Fukugita:1986hr} or through the Affleck-Dine mechanism \cite{Affleck:1984fy}. We intend to return to these open questions in a future work.

\section{Thermal metastability and phase transitions in the early universe}

In the previous section, we proved that the thermal corrections do not completely destabilize the vacuum, although the theory predicts a decompactification temperature above which the fate of the effective theory is at stake. However, it would be interesting to examine the possibility of the thermal metastability of the vacuum. This aspect of the string landscape has raised a fruitful discussion regarding the generation of gravitational waves \cite{Conlon:2025mqt, SanchezGonzalez:2025uco} and a potential interpretation of dark energy as quintessence \cite{Hardy:2019apu, Gomes:2023dat, Cicoli:2023opf}. Previous studies have shown that the lifetime of these vacua is long enough to surpass the age of the universe, either considering Colemen-De Lucia (CdL) \cite{Coleman:1977py, Coleman:1980aw} or HM instanton solutions \cite{Hawking:1981fz}. This fact can be attributed to the large enough barrier that separates the false vacuum and the true vacuum, which in the most cases is the true vacuum at infinity. The indirect effect of the D-term uplift is that the uplifted vacuum flattens faster than the barrier shrinks, thereby preventing the formation of CdL instantons that would potentially trigger strong first-order transitions in the early universe. 
\par Having this in mind, in this section, we try to probe the temperature that signifies the emergence of HM instanton solutions in the presence of finite temperature corrections. These events are of particular importance due to their classical picture of the modulus jumping to the top of the potential and rolling classically towards the real vacuum. In order to do so, we first examine whether there exists a critical temperature $T_{crit}<T_{max}$ at which the reduction of the barrier height outpaces the flattening of the vacuum. As a result, the tunneling channels open before the decompactification temperature is reached, making moduli tunneling the preferred decay mode. The analysis in the previous section comes in handy, especially the definition of the $T_{max}$, where as the temperature increases, we expect in some region the two extrema (minimum and maximum) to coincide. The approximate formula for the minimum we found before effectively describes a small vicinity of the temperatures, away from the boundaries of the Lambert function. As the temperature increases, the prefactor of the $\mathcal{O} (T^2)$ term should decrease to reach the maximum point. Thus, a critical temperature $T_{crit}$ can be found by demanding the following inequality:
\begin{align}
    \dfrac{dB}{dT} = \frac{d \Delta V}{dT},\quad \Delta V =& V(t_{max},T)-V(t_{min},T),\; B: V(t_{max},T)\quad \text{barrier height},\\
    &|\dfrac{dB}{dT}| > |\dfrac{dV(t_{min},T)}{dT}|~.\label{criterion}
\end{align}
Now, as discussed before, the parameter that characterizes the critical temperature is the $n$ parameter in the definition of a minimal value.
\begin{align}
    &t_{min}(n) \cong \frac{-12 \sqrt{6} \eta\; W_{0}\left(-\frac{e^{\frac{13}{3}-\frac{\xi }{4 \eta }} \left(\sqrt{\frac{n}{\pi }}\mathcal{W}_0 \sqrt{-\mathcal{C}_1 g_s T^4 }-2 d\right)}{9 \eta  \mathcal{W}_0^2 g_s}\right)+52 \sqrt{6} \eta -3 \sqrt{6} \xi }{36 \eta },\label{min}\\
    &t_{max}(n) \cong \frac{-12 \sqrt{6} \eta\; W_{-1}\left(-\frac{e^{\frac{13}{3}-\frac{\xi }{4 \eta }} \left(\sqrt{\frac{n-1}{\pi }}\mathcal{W}_0 \sqrt{-\mathcal{C}_1 g_s T^4 }-2 d\right)}{9 \eta  \mathcal{W}_0^2 g_s}\right)+52 \sqrt{6} \eta -3 \sqrt{6} \xi }{36 \eta }~.\label{max}
\end{align}
So, the critical temperature $T_{crit}$ is bounded by the critical value of $n$, which can be observed by the following figures.
\begin{equation}
    2d + 9g_s \eta \; \mathcal{W}_0^2 e^{+\frac{\xi }{4 \eta }-\frac{16}{3}} = f(d,\mathcal{W}_0) \leq \sqrt{\frac{n}{\pi}} \sqrt{-\mathcal{C}_1 g_s} \mathcal{W}_0 T_{crit}^2 = h(n,T)~.\label{critT}
\end{equation}
This bound can be perceived as a deviation in the prefactor of the term $\mathcal{O} (T^2)$, which accounts for the shift of the vacuum. Now, the reason why this critical temperature $T_{\text{crit}}$ satisfies inequality~\eqref{criterion} is that the solution must be pushed sufficiently close to the boundary of the Lambert $W_{-1}$ branch, where thermal effects can compete with (and possibly overcome) the uplift. The D-term uplift scales differently from the thermal uplift and, in particular, it cannot reduce the barrier width efficiently enough to trigger tunneling. In Figure 4., we display the critical value of the parameter $n$ within the minimal range considered, which corresponds to the critical temperature $T_{\text{crit}}$. For temperatures above $T_{\text{crit}}$, the inequality~\eqref{criterion} is satisfied, indicating that in this regime Hawking-Moss instantons dominate over CDL instantons.  
\begin{figure}[H]
    \centering
    \includegraphics[scale=1]{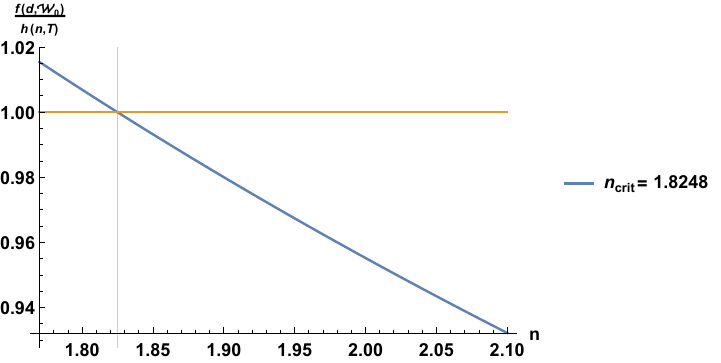}
    \caption{Ratio of functions in inequality \eqref{critT} characterizing the critical thermal effects on the vacuum. The critical temperature is given $T(n_{crit})=0.011$ and a slightly increased $\lambda=6\times 10^{-3}$ is needed, in order to satisfy the criterion \eqref{criterion}. the rest of the parameters are given by Table 1.}
\end{figure}
\noindent The above observation leads us to consider the implications of the HM tunneling of the current vacuum. The criterion that has to be satisfied for the emergence of the Hawking-Moss instanton, is described by the following equation \cite{Jensen:1983ac, Balek:2003uu}:
\begin{equation}
    H^2_c = \dfrac{V''(t_{max})}{4}+\dfrac{\Delta V}{3},\quad H_{\pm} = \sqrt{\dfrac{V(t_{min},t_{max})}{3}},
\end{equation}
where $H$ is the Hubble parameter and $H_c$ denotes the critical Hubble parameter that discriminates between the different instanton solutions. Our finite temperature induced vacuum is almost flat, thus $\Delta V$ term can be safely neglected and, then, the criterion reads as \cite{Jensen:1983ac, Balek:2003uu}:
\begin{equation}
    \dfrac{H_c^2}{H^2_-} = \dfrac{3}{4} \dfrac{V''(t_{max})}{V(t_{min})}~.\label{HW}
\end{equation}
For $H_- > H_c$, the CdL instantons cannot be justified, which leads to a classical description of the field jumping above the barrier. The Euclidean action of the instanton solution in a symmetric $O(4)$ symmetric background can be written as:
\begin{align}
    S_E(t) = -2\pi^2\int d\rho \dfrac{1}{H_{\pm}^3}\sin^3(H_{\pm} \rho) V(t) = -\dfrac{24\pi^2}{V(t_{\pm})}, \quad ds^2 = d\rho^2 + \chi(\rho)^2d\Omega^2,
\end{align}
where $\rho$ is the time variable and $\chi$ is the scale factor of the sphere. Returning to the criterion of equation \eqref{HW}, we compute it for the parameters of Table 3. except ($T=0.011,\; \lambda=6\times 10^{-3}$), where the minima of the potential are described by \eqref{min},\eqref{max}.  
\begin{align}
    \dfrac{H_c^2}{H^2_-}|_{dS} \sim 0.88~.
\end{align}
The tunneling coefficient for the process can be numerically expressed by:
\begin{equation}
    \Gamma = A e^{-B},\quad  B|_{dS} = -24\pi^2 (\dfrac{1}{V(t_{min})}-\dfrac{1}{V(t_{max})})|_{dS} \sim 10^{11}~.
\end{equation}
It is evident that the finite temperature corrections cannot destabilize the vacuum. So, the volume direction $t$ which is the lightest direction does not provide any new physics. Thus, we expect that as the temperature decreases, a classical continuous second-order–like phase transition will occur towards the $T=0$ vacuum. More interestingly, let us plot the potential along the thermalization direction, which is the $\tau_1$ direction. The true vacuum at $T=0$ lies at $\tau_1 =7$, where upon the inclusion of finite temperature corrections, the slope of the potential pushes the vacuum to large $\tau_1$. However, finite temperature corrections create a barrier (Figure 5.) between a new thermal AdS and the thermal metastable dS minimum described above. This fact opens up a new possibility to explain the metastability of the dS vacua in string compactifications, since plausible first-order phase transitions could emerge as the universe cools down.
\begin{figure}[H]
    \centering
    \includegraphics[scale=0.7]{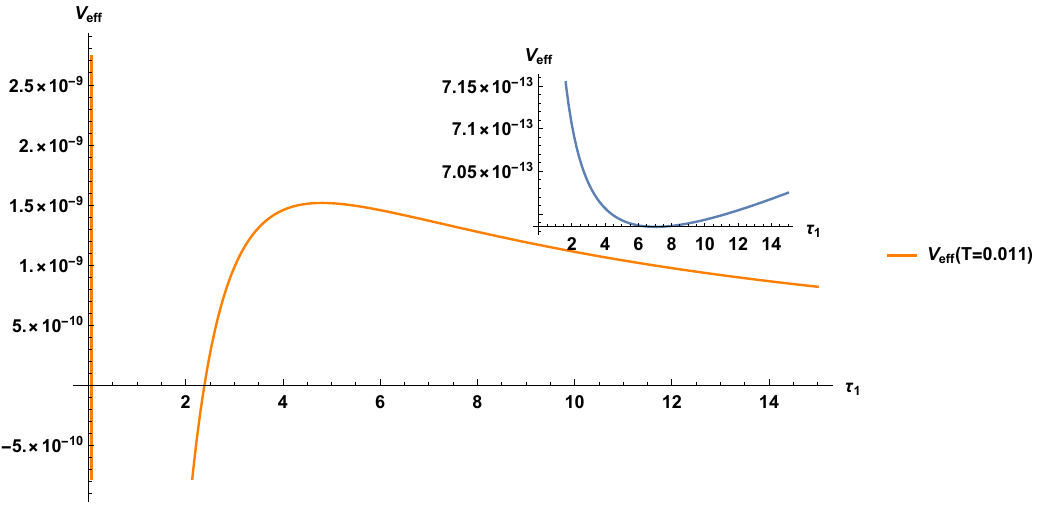}
    \caption{Plot at the potential \eqref{Vtot} (orange curve) with $T\neq 0$ along the $\tau_1$ direction, while the blue curve shows the minimum at $T=0$.}
\end{figure}
\noindent Consequently, we could denote two separate cases that associate the metastability of the vacuum with the reheating scale. 
\begin{itemize}
    \item \underline{$H_{inf}<T_{crit}$}: If the inflation scale is very high and the reheating process is instantaneous, then the finite temperature corrections displace the vacuum very far away from the $T=0$ minimum, where the volume direction is safe from tunneling. In this case, a continuous second-order–like phase transition will occur towards the return of the moduli at the true vacuum at $T=0$. The generation of gravitational waves in this picture cannot be attributed to a phase transition, but the possibility of an oscillon condensation is very likely due to the oscillation of the field in the thermalized vacuum \cite{Antusch:2017flz, Gallego:2020vbe}. Earlier studies discussed the emergence of these instabilities of the thermal bath, which lead to low frequency gravitational waves closer to the currently observable range.
    \item \underline{$H_{inf}>T_{crit}$}: Finite temperature corrections again displace the vacuum away from the $T=0$ minimum, but the potential generates a second AdS vacuum at small $\tau_1$ values. Thus, in this region of parameter space, first-order phase transitions or thermal bounces could lead to an interesting cosmological history, where the fate of the universe will result in a big crunch. However, the supergravity approximation and the loop corrections need to be further understood, in order to make definite statements about the metastability nature of the proposed vacuum structure.
\end{itemize}
\noindent Since the $\tau_1$ modulus is the heaviest field, the transition should be treated as a multi-field process, while the bounces can be treated as thermal bubbles \cite{Hiramatsu:2014uta, Lee:1987qc}. A further indication that first-order thermal transitions could arise in our setup comes from recent developments in the study of vacuum transitions in the string landscape \footnote{Similar analysis showed some indications for high energy phase transitions in heterotic string theory \cite{Kounnas:2007vd, Kounnas:2007hb}. A more recent study correlates the Festina Lente Swampland bound with the thermal effects \cite{Venken:2023hfa}.}. In particular, the Hamiltonian analysis of \cite{Cespedes:2023jdk, Cespedes:2020xpn, Pasquarella:2022ibb} suggests that the standard Euclidean CdL picture may underestimate the effective connectivity between vacua. The Euclidean approach typically treats de~Sitter, Minkowski, and AdS vacua as pure states with vanishing entropy, which tends to suppress or entirely forbid certain transitions. By contrast, in the Hamiltonian formalism these vacua behave as mixed states with non-zero entropy, and the corresponding transition rates need not vanish. As argued in \cite{Cespedes:2023jdk}, both up- and down-tunneling processes may become allowed once entropic contributions are included, and many transitions can be viewed as limits of transitions involving small black holes in the $M\!\to\!0$ limit. From this perspective, the thermal bath that displaces the $\tau_1$ modulus in our model may play a role analogous to the entropy contribution of a small black hole: it effectively promotes the finite-temperature de Sitter configuration to a mixed state with an increased number of accessible microstates. This raises the possibility that the thermal metastable vacuum generated at large $\tau_1$ could undergo transitions into the competing AdS region even in situations where a naive Euclidean CDL analysis would indicate extreme suppression. Although such transitions remain model-dependent, the entropic reasoning of \cite{Cespedes:2023jdk} suggests that they need not be entirely negligible in a finite-temperature cosmological environment.\par
In our setup, the thermal corrections generate an additional AdS extremum at small $\tau_1$, separated from the displaced de~Sitter minimum by a temperature-dependent barrier. The arguments of \cite{Cespedes:2023jdk} may offer qualitative support for this interpretation: since the thermal de~Sitter phase carries an entropy proportional to $H^{-2}(T)$, detailed balance may allow transitions from the high-entropy thermal configuration toward the lower-entropy AdS minimum, including processes that would be highly suppressed at zero temperature. Furthermore, the possible presence of primordial black holes in the early universe could enhance such effects, as their microstates modify tunneling amplitudes in a similar manner as identified in \cite{Cespedes:2023jdk}. The generation of primordial black holes at the early universe have been studied extensively in the past \cite{Kaloper:2004yj, ShamsEsHaghi:2022azq, Sandick:2021gew, Kawasaki:2018daf}, where their presence could have implications on baryogenesis as well as on the their role as catalyst for phase transitions. Therefore, the thermal AdS vacuum emerging in the $\tau_1$ direction may be viewed as potentially accessible through entropy-assisted vacuum transitions, consistent with the broader quantum-gravitational picture emerging from the Hamiltonian formulation. We do not state that our thermal entropy of the dS vacuum is a direct analogue of the horizon entropy, but we tried to schematically provide a conceptual motivation.\par
Taken together, these considerations point towards a scenario in which the two cases discussed above, $H_{\rm inf} < T_{\rm crit}$ and $H_{\rm inf} > T_{\rm crit}$, might correspond to different possible cosmological histories within the landscape. In the first case, the thermal displacement of $\tau_1$ could remain in a regime where entropic effects are insufficient to overcome the stabilizing potential, leading the modulus to roll back smoothly to the $T=0$ vacuum. In the second case, however, the thermal state may enter the region where entropy-assisted transitions of the type described in \cite{Cespedes:2023jdk} could become relevant, opening a possible pathway for bubble nucleation into the emergent thermal AdS vacuum and initiating a big-crunch cosmology. Such a mechanism would provide a landscape-motivated explanation for how different reheating scales might lead to different cosmological outcomes, with the metastability of de~Sitter vacua influenced not only by the zero-temperature potential but also by their thermal and entropic properties.

\section{Conclusions}

In this work, we have analyzed the finite-temperature cosmological dynamics of a perturbatively stabilized LVS compactifications including logarithmic loop corrections, winding effects, and $F^{4}$ higher-derivative contributions. We demonstrated that one of the transverse Kähler moduli can thermalize after reheating, inducing a temperature-dependent shift of the vacuum and generating a metastable de Sitter configuration at high energies. Within this framework, we derived analytic expressions for the thermally shifted minima and the corresponding decompactification temperature $T_{\max}$, explicitly relating these quantities to the microscopic parameters of the model. Moreover, we demonstrated that finite temperature effects on the stabilization process emerge through the winding-loop sector, where the corresponding factor $\mathcal{C}_1$ parametrizes the aforementioned contributions to the effective theory. This opens up questions on the dynamics of the complex structure moduli sector at high energies, where the position of the branes and the flux background will specify the order of magnitude for the coefficients $\mathcal{C}_i$. Our analysis shows that, although the volume modulus remains safely stabilized for all temperatures below $T_{\max}$, thermal corrections in the transverse direction may create an additional AdS extremum separated by a shrinking barrier. This opens the possibility that, for temperatures above a critical value $T_{\rm crit}$, plausible first-order transitions or thermal bounces could temporarily alter the cosmological evolution before the system relaxes back to the $T=0$ vacuum. Furthermore, we examined the temperatures where the thermalized moduli will decay, which lead to the observation that their energy density cannot dominate the universe and so moduli-dominated eras are disfavored for high scale reheating energies in the current framework.\par
We also discussed how these features may connect with recent developments in the Hamiltonian description of vacuum transitions. In particular, the work of \cite{Cespedes:2023jdk} suggests that the entropic nature of finite-temperature (and black-hole–seeded) configurations can modify the effective connectivity of the string landscape, allowing transitions that are strongly suppressed in the Euclidean CDL framework. While our model remains within the controlled supergravity regime, the appearance of a thermally generated AdS vacuum provides a concrete setting in which such entropy-assisted transitions could, in principle, play a role. Future directions include a full multi-field bounce computation in the field space, an exploration of whether black-hole–induced channels enhance or suppress the plausible transitions identified here, and a study of how different reheating scales may select different cosmological histories within the landscape taking into account the flux background with its implications on dilaton and complex structure moduli. These questions could help clarify whether thermal metastability in perturbative LVS might contribute to a broader understanding of vacuum selection and the fate of de Sitter solutions in string theory.

\section*{Acknowledgments}
We are thankful to Emeritus Professor George Leontaris for a crucial read and comments on the final draft of this paper. This research work is funded by the Research Committee of the University of Ioannina under the fellowship program for postdoctoral studies.


\begin{thebibliography}{99}

\bibitem{Derendinger:2004jn}
J.~P.~Derendinger, C.~Kounnas, P.~M.~Petropoulos and F.~Zwirner,
Nucl. Phys. B \textbf{715} (2005), 211-233
doi:10.1016/j.nuclphysb.2005.02.038
[arXiv:hep-th/0411276 [hep-th]].

\bibitem{Villadoro:2005cu}
G.~Villadoro and F.~Zwirner,
JHEP \textbf{06} (2005), 047
doi:10.1088/1126-6708/2005/06/047
[arXiv:hep-th/0503169 [hep-th]].

\bibitem{DeWolfe:2005uu}
O.~DeWolfe, A.~Giryavets, S.~Kachru and W.~Taylor,
JHEP \textbf{07} (2005), 066
doi:10.1088/1126-6708/2005/07/066
[arXiv:hep-th/0505160 [hep-th]].

\bibitem{Giddings:2001yu}
S.~B.~Giddings, S.~Kachru and J.~Polchinski,
Phys. Rev. D \textbf{66} (2002), 106006
doi:10.1103/PhysRevD.66.106006
[arXiv:hep-th/0105097 [hep-th]].

\bibitem{hep-th/0301240}
S.~Kachru, R.~Kallosh, A.~D.~Linde and S.~P.~Trivedi,
Phys. Rev. D \textbf{68} (2003), 046005
doi:10.1103/PhysRevD.68.046005
[arXiv:hep-th/0301240 [hep-th]].

\bibitem{hep-th/0502058}
V.~Balasubramanian, P.~Berglund, J.~P.~Conlon and F.~Quevedo,
JHEP \textbf{03} (2005), 007
doi:10.1088/1126-6708/2005/03/007
[arXiv:hep-th/0502058 [hep-th]].

\bibitem{arXiv:0805.1029}
M.~Cicoli, J.~P.~Conlon and F.~Quevedo,
JHEP \textbf{10} (2008), 105
doi:10.1088/1126-6708/2008/10/105
[arXiv:0805.1029 [hep-th]].

\bibitem{Vafa:2005ui}
C.~Vafa,
[arXiv:hep-th/0509212 [hep-th]].

\bibitem{Ooguri:2006in}
H.~Ooguri and C.~Vafa,
Nucl. Phys. B \textbf{766} (2007), 21-33
doi:10.1016/j.nuclphysb.2006.10.033
[arXiv:hep-th/0605264 [hep-th]].

\bibitem{Grana:2021zvf}
M.~Gra{\~n}a and A.~Herr{\'a}ez,
Universe \textbf{7} (2021) no.8, 273
doi:10.3390/universe7080273
[arXiv:2107.00087 [hep-th]].

\bibitem{hep-th/0411011}
R.~Kallosh and A.~D.~Linde,
JHEP \textbf{12} (2004), 004
doi:10.1088/1126-6708/2004/12/004
[arXiv:hep-th/0411011 [hep-th]].

\bibitem{Binetruy:1984yx}
P.~Binetruy and M.~K.~Gaillard,
Phys. Rev. D \textbf{32} (1985), 931-937
doi:10.1103/PhysRevD.32.931

\bibitem{Binetruy:1984wy}
P.~Binetruy and M.~K.~Gaillard,
Nucl. Phys. B \textbf{254} (1985), 388-424
doi:10.1016/0550-3213(85)90225-1

\bibitem{Buchmuller:2004tz}
W.~Buchmuller, K.~Hamaguchi, O.~Lebedev and M.~Ratz,
JCAP \textbf{01} (2005), 004
doi:10.1088/1475-7516/2005/01/004
[arXiv:hep-th/0411109 [hep-th]].

\bibitem{Anguelova:2007at}
L.~Anguelova, R.~Ricci and S.~Thomas,
Phys. Rev. D \textbf{77} (2008), 025036
doi:10.1103/PhysRevD.77.025036
[arXiv:hep-th/0702168 [hep-th]].

\bibitem{Fischler:2006xh}
W.~Fischler, V.~Kaplunovsky, C.~Krishnan, L.~Mannelli and M.~A.~C.~Torres,
JHEP \textbf{03} (2007), 107
doi:10.1088/1126-6708/2007/03/107
[arXiv:hep-th/0611018 [hep-th]].

\bibitem{Gross:1982cv}
D.~J.~Gross, M.~J.~Perry and L.~G.~Yaffe,
Phys. Rev. D \textbf{25} (1982), 330-355
doi:10.1103/PhysRevD.25.330

\bibitem{Linde:1978px}
A.~D.~Linde,
Rept. Prog. Phys. \textbf{42} (1979), 389
doi:10.1088/0034-4885/42/3/001

\bibitem{Linde:1980tt}
A.~D.~Linde,
Phys. Lett. B \textbf{100} (1981), 37-40
doi:10.1016/0370-2693(81)90281-1

\bibitem{Linde:1981zj}
A.~D.~Linde,
Nucl. Phys. B \textbf{216} (1983), 421
[erratum: Nucl. Phys. B \textbf{223} (1983), 544]
doi:10.1016/0550-3213(83)90072-X

\bibitem{Linde:1982tg}
A.~D.~Linde,
Phys. Lett. B \textbf{116} (1982), 340-342
doi:10.1016/0370-2693(82)90294-5

\bibitem{hep-th/0611006}
N.~J.~Craig, P.~J.~Fox and J.~G.~Wacker,
Phys. Rev. D \textbf{75} (2007), 085006
doi:10.1103/PhysRevD.75.085006
[arXiv:hep-th/0611006 [hep-th]].

\bibitem{hep-th/0610334}
S.~A.~Abel, C.~S.~Chu, J.~Jaeckel and V.~V.~Khoze,
JHEP \textbf{01} (2007), 089
doi:10.1088/1126-6708/2007/01/089
[arXiv:hep-th/0610334 [hep-th]].

\bibitem{hep-th/0611018}
W.~Fischler, V.~Kaplunovsky, C.~Krishnan, L.~Mannelli and M.~A.~C.~Torres,
JHEP \textbf{03} (2007), 107
doi:10.1088/1126-6708/2007/03/107
[arXiv:hep-th/0611018 [hep-th]].

\bibitem{hep-th/0702168}
L.~Anguelova, R.~Ricci and S.~Thomas,
Phys. Rev. D \textbf{77} (2008), 025036
doi:10.1103/PhysRevD.77.025036
[arXiv:hep-th/0702168 [hep-th]].

\bibitem{Papineau:2008xf}
C.~Papineau,
JHEP \textbf{05} (2008), 068
doi:10.1088/1126-6708/2008/05/068
[arXiv:0802.1861 [hep-th]].

\bibitem{hep-th/0404168}
W.~Buchmuller, K.~Hamaguchi, O.~Lebedev and M.~Ratz,
Nucl. Phys. B \textbf{699} (2004), 292-308
doi:10.1016/j.nuclphysb.2004.08.031
[arXiv:hep-th/0404168 [hep-th]].

\bibitem{Leontaris:2022rzj}
G.~K.~Leontaris and P.~Shukla,
JHEP \textbf{07} (2022), 047
doi:10.1007/JHEP07(2022)047
[arXiv:2203.03362 [hep-th]].

\bibitem{Basiouris:2020jgp}
V.~Basiouris and G.~K.~Leontaris,
Phys. Lett. B \textbf{810} (2020), 135809
doi:10.1016/j.physletb.2020.135809
[arXiv:2007.15423 [hep-th]].

\bibitem{Basiouris:2021sdf}
V.~Basiouris and G.~K.~Leontaris,
Fortsch. Phys. \textbf{70} (2022) no.2-3, 2100181
doi:10.1002/prop.202100181
[arXiv:2109.08421 [hep-th]].

\bibitem{Bera:2024zsk}
S.~Bera, D.~Chakraborty, G.~K.~Leontaris and P.~Shukla,
JCAP \textbf{09} (2024), 004
doi:10.1088/1475-7516/2024/09/004
[arXiv:2405.06738 [hep-th]].

\bibitem{Hai:2025wvs}
M.~Hai, A.~R.~Kamal, N.~F.~Shamma and M.~S.~J.~Shuvo,
[arXiv:2506.08083 [hep-th]].

\bibitem{Chakraborty:2025wqn}
D.~Chakraborty, M.~Hai, S.~T.~Jahan, A.~R.~Kamal and M.~S.~J.~Shuvo,
[arXiv:2511.19610 [hep-th]].

\bibitem{Hawking:1981fz}
S.~W.~Hawking and I.~G.~Moss,
Phys. Lett. B \textbf{110} (1982), 35-38
doi:10.1016/0370-2693(82)90946-7

\bibitem{hep-th/0204254}
K.~Becker, M.~Becker, M.~Haack and J.~Louis,
JHEP \textbf{06} (2002), 060
doi:10.1088/1126-6708/2002/06/060
[arXiv:hep-th/0204254 [hep-th]].

\bibitem{Antoniadis:2018hqy}
I.~Antoniadis, Y.~Chen and G.~K.~Leontaris,
Eur. Phys. J. C \textbf{78} (2018) no.9, 766
doi:10.1140/epjc/s10052-018-6248-4
[arXiv:1803.08941 [hep-th]].

\bibitem{Antoniadis:2019rkh}
I.~Antoniadis, Y.~Chen and G.~K.~Leontaris,
JHEP \textbf{01} (2020), 149
doi:10.1007/JHEP01(2020)149
[arXiv:1909.10525 [hep-th]].

\bibitem{hep-th/9906070}
S.~Gukov, C.~Vafa and E.~Witten,
Nucl. Phys. B \textbf{584} (2000), 69-108
[erratum: Nucl. Phys. B \textbf{608} (2001), 477-478]
doi:10.1016/S0550-3213(00)00373-4
[arXiv:hep-th/9906070 [hep-th]].

\bibitem{Kreuzer:2000xy}
M.~Kreuzer and H.~Skarke,
Adv. Theor. Math. Phys. \textbf{4} (2000), 1209-1230
doi:10.4310/ATMP.2000.v4.n6.a2
[arXiv:hep-th/0002240 [hep-th]].

\bibitem{Altman:2014bfa}
R.~Altman, J.~Gray, Y.~H.~He, V.~Jejjala and B.~D.~Nelson,
JHEP \textbf{02} (2015), 158
doi:10.1007/JHEP02(2015)158
[arXiv:1411.1418 [hep-th]].

\bibitem{Blumenhagen:2011xn}
R.~Blumenhagen, B.~Jurke and T.~Rahn,
Adv. High Energy Phys. \textbf{2011} (2011), 152749
doi:10.1155/2011/152749
[arXiv:1104.1187 [hep-th]].

\bibitem{Blumenhagen:2010pv}
R.~Blumenhagen, B.~Jurke, T.~Rahn and H.~Roschy,
J. Math. Phys. \textbf{51} (2010), 103525
doi:10.1063/1.3501132
[arXiv:1003.5217 [hep-th]].

\bibitem{hep-th/0508043}
M.~Berg, M.~Haack and B.~Kors,
JHEP \textbf{11} (2005), 030
doi:10.1088/1126-6708/2005/11/030
[arXiv:hep-th/0508043 [hep-th]].

\bibitem{Cicoli:2007xp}
M.~Cicoli, J.~P.~Conlon and F.~Quevedo,
JHEP \textbf{01} (2008), 052
doi:10.1088/1126-6708/2008/01/052
[arXiv:0708.1873 [hep-th]].

\bibitem{Berg:2004ek}
M.~Berg, M.~Haack and B.~Kors,
Phys. Rev. D \textbf{71} (2005), 026005
doi:10.1103/PhysRevD.71.026005
[arXiv:hep-th/0404087 [hep-th]].

\bibitem{Berg:2005yu}
M.~Berg, M.~Haack and B.~Kors,
Phys. Rev. Lett. \textbf{96} (2006), 021601
doi:10.1103/PhysRevLett.96.021601
[arXiv:hep-th/0508171 [hep-th]].

\bibitem{Ciupke:2015msa}
D.~Ciupke, J.~Louis and A.~Westphal,
JHEP \textbf{10} (2015), 094
doi:10.1007/JHEP10(2015)094
[arXiv:1505.03092 [hep-th]].

\bibitem{Dolan:1973qd}
L.~Dolan and R.~Jackiw,
Phys. Rev. D \textbf{9} (1974), 3320-3341
doi:10.1103/PhysRevD.9.3320

\bibitem{Jackiw:1974cv}
R.~Jackiw,
Phys. Rev. D \textbf{9} (1974), 1686
doi:10.1103/PhysRevD.9.1686

\bibitem{Anguelova:2007ex}
L.~Anguelova and V.~Calo,
Nucl. Phys. B \textbf{801} (2008), 45-69
doi:10.1016/j.nuclphysb.2008.04.009
[arXiv:0708.4159 [hep-th]].

\bibitem{Antoniadis:2018ngr}
I.~Antoniadis, Y.~Chen and G.~K.~Leontaris,
Int. J. Mod. Phys. A \textbf{34} (2019) no.08, 1950042
doi:10.1142/S0217751X19500428
[arXiv:1810.05060 [hep-th]].

\bibitem{Cicoli:2010ha}
M.~Cicoli and A.~Mazumdar,
JCAP \textbf{09} (2010), 025
doi:10.1088/1475-7516/2010/09/025
[arXiv:1005.5076 [hep-th]].

\bibitem{hep-th/0610129}
J.~P.~Conlon, S.~S.~Abdussalam, F.~Quevedo and K.~Suruliz,
JHEP \textbf{01} (2007), 032
doi:10.1088/1126-6708/2007/01/032
[arXiv:hep-th/0610129 [hep-th]].

\bibitem{Anguelova:2009ht}
L.~Anguelova, V.~Calo and M.~Cicoli,
JCAP \textbf{10} (2009), 025
doi:10.1088/1475-7516/2009/10/025
[arXiv:0904.0051 [hep-th]].

\bibitem{Burgess:2003ic}
C.~P.~Burgess, R.~Kallosh and F.~Quevedo,
JHEP \textbf{10} (2003), 056
doi:10.1088/1126-6708/2003/10/056
[arXiv:hep-th/0309187 [hep-th]].

\bibitem{Haack:2006cy}
M.~Haack, D.~Krefl, D.~Lust, A.~Van Proeyen and M.~Zagermann,
JHEP \textbf{01} (2007), 078
doi:10.1088/1126-6708/2007/01/078
[arXiv:hep-th/0609211 [hep-th]].

\bibitem{Achucarro:2006zf}
A.~Achucarro, B.~de Carlos, J.~A.~Casas and L.~Doplicher,
JHEP \textbf{06} (2006), 014
doi:10.1088/1126-6708/2006/06/014
[arXiv:hep-th/0601190 [hep-th]].

\bibitem{Kapusta:2007xjq}
J.~I.~Kapusta and C.~Gale,
Cambridge University Press, 2007,
ISBN 978-1-009-40196-8, 978-1-009-40195-1, 978-1-009-40198-2
doi:10.1017/9781009401968

\bibitem{hep-th/9502077}
V.~Kaplunovsky and J.~Louis,
Nucl. Phys. B \textbf{444} (1995), 191-244
doi:10.1016/0550-3213(95)00172-O
[arXiv:hep-th/9502077 [hep-th]].

\bibitem{Atick:1988si}
J.~J.~Atick and E.~Witten,
Nucl. Phys. B \textbf{310} (1988), 291-334
doi:10.1016/0550-3213(88)90151-4

\bibitem{Catelin-Jullien:2007ewh}
T.~Catelin-Jullien, C.~Kounnas, H.~Partouche and N.~Toumbas,
Nucl. Phys. B \textbf{797} (2008), 137-178
doi:10.1016/j.nuclphysb.2007.12.030
[arXiv:0710.3895 [hep-th]].

\bibitem{Liu:2011nw}
L.~Liu and H.~Partouche,
JHEP \textbf{11} (2012), 079
doi:10.1007/JHEP11(2012)079
[arXiv:1111.7307 [hep-th]].

\bibitem{Baer:2022fou}
H.~Baer, V.~Barger and R.~W.~Deal,
JHEAp \textbf{34} (2022), 40-48
doi:10.1016/j.jheap.2022.04.001
[arXiv:2204.01130 [hep-ph]].

\bibitem{Basiouris:2024buq}
V.~Basiouris,
Phys. Rev. D \textbf{112} (2025) no.7, 075003
doi:10.1103/cngs-jfjb
[arXiv:2411.18737 [hep-ph]].

\bibitem{Cicoli:2022uqa}
M.~Cicoli, K.~Sinha and R.~Wiley Deal,
JHEP \textbf{12} (2022), 068
doi:10.1007/JHEP12(2022)068
[arXiv:2208.01017 [hep-th]].

\bibitem{Ling:2025nlw}
S.~Ling, A.~J.~Long, E.~McDonough and A.~Hayes,
Phys. Rev. D \textbf{112} (2025) no.2, 023550
doi:10.1103/wyc2-fytd
[arXiv:2504.13256 [hep-th]].

\bibitem{WileyDeal:2023trg}
R.~Wiley Deal, K.~Sankharva, K.~Sinha and S.~Watson,
JHEP \textbf{02} (2024), 085
doi:10.1007/JHEP02(2024)085
[arXiv:2308.16242 [hep-ph]].

\bibitem{Allahverdi:2010im}
R.~Allahverdi, B.~Dutta and K.~Sinha,
Phys. Rev. D \textbf{82} (2010), 035004
doi:10.1103/PhysRevD.82.035004
[arXiv:1005.2804 [hep-ph]].

\bibitem{Dutta:2010sg}
B.~Dutta and K.~Sinha,
Phys. Rev. D \textbf{82} (2010), 095003
doi:10.1103/PhysRevD.82.095003
[arXiv:1008.0148 [hep-th]].

\bibitem{Allahverdi:2010rh}
R.~Allahverdi, B.~Dutta and K.~Sinha,
Phys. Rev. D \textbf{83} (2011), 083502
doi:10.1103/PhysRevD.83.083502
[arXiv:1011.1286 [hep-ph]].

\bibitem{Higaki:2012ba}
T.~Higaki, K.~Kamada and F.~Takahashi,
JHEP \textbf{09} (2012), 043
doi:10.1007/JHEP09(2012)043
[arXiv:1207.2771 [hep-ph]].

\bibitem{Garcia:2013bha}
M.~A.~G.~Garcia and K.~A.~Olive,
JCAP \textbf{09} (2013), 007
doi:10.1088/1475-7516/2013/09/007
[arXiv:1306.6119 [hep-ph]].

\bibitem{Fukugita:1986hr}
M.~Fukugita and T.~Yanagida,
Phys. Lett. B \textbf{174} (1986), 45-47
doi:10.1016/0370-2693(86)91126-3

\bibitem{Affleck:1984fy}
I.~Affleck and M.~Dine,
Nucl. Phys. B \textbf{249} (1985), 361-380
doi:10.1016/0550-3213(85)90021-5

\bibitem{Conlon:2025mqt}
J.~P.~Conlon, E.~J.~Copeland, E.~Hardy and N.~S{\'a}nchez Gonz{\'a}lez,
[arXiv:2511.16404 [hep-ph]].

\bibitem{SanchezGonzalez:2025uco}
N.~S{\'a}nchez Gonz{\'a}lez, J.~P.~Conlon, E.~J.~Copeland and E.~Hardy,
JHEP \textbf{10} (2025), 121
doi:10.1007/JHEP10(2025)121
[arXiv:2505.14187 [hep-ph]].

\bibitem{Hardy:2019apu}
E.~Hardy and S.~Parameswaran,
Phys. Rev. D \textbf{101} (2020) no.2, 023503
doi:10.1103/PhysRevD.101.023503
[arXiv:1907.10141 [hep-th]].

\bibitem{Gomes:2023dat}
J.~M.~Gomes, E.~Hardy and S.~Parameswaran,
Phys. Rev. D \textbf{110} (2024) no.2, 2
doi:10.1103/PhysRevD.110.023533
[arXiv:2311.08888 [hep-ph]].

\bibitem{Cicoli:2023opf}
M.~Cicoli, J.~P.~Conlon, A.~Maharana, S.~Parameswaran, F.~Quevedo and I.~Zavala,
Phys. Rept. \textbf{1059} (2024), 1-155
doi:10.1016/j.physrep.2024.01.002
[arXiv:2303.04819 [hep-th]].

\bibitem{Coleman:1977py}
S.~R.~Coleman,
Phys. Rev. D \textbf{15} (1977), 2929-2936
[erratum: Phys. Rev. D \textbf{16} (1977), 1248]
doi:10.1103/PhysRevD.16.1248

\bibitem{Coleman:1980aw}
S.~R.~Coleman and F.~De Luccia,
Phys. Rev. D \textbf{21} (1980), 3305
doi:10.1103/PhysRevD.21.3305

\bibitem{Jensen:1983ac}
L.~G.~Jensen and P.~J.~Steinhardt,
Nucl. Phys. B \textbf{237} (1984), 176-188
doi:10.1016/0550-3213(84)90021-X

\bibitem{Balek:2003uu}
V.~Balek and M.~Demetrian,
Phys. Rev. D \textbf{69} (2004), 063518
doi:10.1103/PhysRevD.69.063518
[arXiv:gr-qc/0311040 [gr-qc]].

\bibitem{Antusch:2017flz}
S.~Antusch, F.~Cefala, S.~Krippendorf, F.~Muia, S.~Orani and F.~Quevedo,
JHEP \textbf{01} (2018), 083
doi:10.1007/JHEP01(2018)083
[arXiv:1708.08922 [hep-th]].

\bibitem{Gallego:2020vbe}
D.~Gallego,
JCAP \textbf{09} (2020), 033
doi:10.1088/1475-7516/2020/09/033
[arXiv:2005.03939 [hep-th]].

\bibitem{Hiramatsu:2014uta}
T.~Hiramatsu, Y.~Miyamoto and J.~Yokoyama,
JCAP \textbf{03} (2015), 024
doi:10.1088/1475-7516/2015/03/024
[arXiv:1412.7814 [hep-ph]].

\bibitem{Lee:1987qc}
K.~M.~Lee and E.~J.~Weinberg,
Phys. Rev. D \textbf{36} (1987), 1088
doi:10.1103/PhysRevD.36.1088

\bibitem{Kounnas:2007vd}
C.~Kounnas and H.~Partouche,
Nucl. Phys. B \textbf{793} (2008), 131-159
doi:10.1016/j.nuclphysb.2007.10.008
[arXiv:0705.3206 [hep-th]].

\bibitem{Kounnas:2007hb}
C.~Kounnas and H.~Partouche,
Nucl. Phys. B \textbf{795} (2008), 334-360
doi:10.1016/j.nuclphysb.2007.11.020
[arXiv:0706.0728 [hep-th]].

\bibitem{Venken:2023hfa}
V.~Venken,
JHEP \textbf{02} (2024), 114
doi:10.1007/JHEP02(2024)114
[arXiv:2311.04955 [hep-th]].

\bibitem{Cespedes:2023jdk}
S.~Cespedes, S.~de Alwis, F.~Muia and F.~Quevedo,
Phys. Rev. D \textbf{109} (2024) no.10, 105027
doi:10.1103/PhysRevD.109.105027
[arXiv:2307.13614 [hep-th]].

\bibitem{Cespedes:2020xpn}
S.~Cespedes, S.~P.~de Alwis, F.~Muia and F.~Quevedo,
Phys. Rev. D \textbf{104} (2021) no.2, 026013
doi:10.1103/PhysRevD.104.026013
[arXiv:2011.13936 [hep-th]].

\bibitem{Pasquarella:2022ibb}
V.~Pasquarella and F.~Quevedo,
JHEP \textbf{05} (2023), 192
doi:10.1007/JHEP05(2023)192
[arXiv:2211.07664 [hep-th]].

\bibitem{Kaloper:2004yj}
N.~Kaloper, J.~Rahmfeld and L.~Sorbo,
Phys. Lett. B \textbf{606} (2005), 234-244
doi:10.1016/j.physletb.2004.11.083
[arXiv:hep-th/0409226 [hep-th]].

\bibitem{ShamsEsHaghi:2022azq}
B.~Shams Es Haghi,
Phys. Rev. D \textbf{107} (2023) no.8, 083507
doi:10.1103/PhysRevD.107.083507
[arXiv:2212.11308 [hep-ph]].

\bibitem{Sandick:2021gew}
P.~Sandick, B.~S.~Es Haghi and K.~Sinha,
Phys. Rev. D \textbf{104} (2021) no.8, 083523
doi:10.1103/PhysRevD.104.083523
[arXiv:2108.08329 [astro-ph.CO]].

\bibitem{Kawasaki:2018daf}
M.~Kawasaki and V.~Takhistov,
Phys. Rev. D \textbf{98} (2018) no.12, 123514
doi:10.1103/PhysRevD.98.123514
[arXiv:1810.02547 [hep-th]].




\end{thebibliography}
\end{document}